\documentclass[aps,prb,reprint,superscriptaddress]{revtex4-2}

\usepackage[english]{babel}
\usepackage{graphicx}
\graphicspath{ {figs/} }
\usepackage{bm}
\usepackage{amsmath}
\usepackage{amsfonts}
\usepackage{soul}
\usepackage{xcolor}

\begin{document}


\title{Tensor Network Based Finite-Size Scaling for Two-Dimensional Classical Models}



\author{Ching-Yu Huang}
\affiliation{ Department of Applied Physics, Tunghai University, Taichung 40704, Taiwan}

\author{Sing-Hong Chan}
\affiliation{ Department of Physics, National Tsing Hua University, Hsinchu 30013, Taiwan}

\author{Ying-Jer Kao }
\email{yjkao@phys.ntu.edu.tw}
\affiliation{ Department of Physics, National Taiwan University, Taipei 10607, Taiwan}
\affiliation{Center for Quantum Science and Technology,  National Taiwan University, Taipei 10607, Taiwan }
\affiliation{National Center for High-Performance Computing, Hsinchu City 30076, Taiwan}

\author{Pochung Chen}
\email{pcchen@phys.nthu.edu.tw}
\affiliation{ Department of Physics, National Tsing Hua University, Hsinchu 30013, Taiwan}
\affiliation{ Physics Division, National Center for Theoretical Sciences, Taipei 10617, Taiwan}
\affiliation{ Frontier Center for Theory and Computation, National Tsing Hua University, Hsinchu 30013, Taiwan}


\date{\today}

\begin{abstract}
  We propose a scheme to perform tensor network based finite-size scaling analysis for two-dimensional classical models.
  In the tensor network representation of the partition function,  we use higher-order tensor renormalization group (HOTRG) method to coarse grain the weight tensor.
  The renormalized tensor is then used to construct the approximated transfer matrix of an infinite strip of finite width.
  By diagonalizing the transfer matrix we obtain the correlation length, the magnetization, and the energy density 
  which are used in finite-size scaling analysis to determine the critical temperature and the critical exponents.
  As a benchmark we study the two-dimensional classical Ising model. 
  We show that the critical temperature and the critical exponents can be accurately determined. 
  With HOTRG bond dimension $D=70$, the absolute errors of the critical temperature $T_c$ and the critical exponent $\nu$, $\beta$
  are at the order of $10^{-7}, 10^{-5}$, $10^{-4}$ respectively.
  Furthermore, the results can be systematically improved by increasing the bond dimension of the HOTRG method.
  Finally, we study the length scale induced by the finite cut-off in bond dimension and elucidate its physical meaning in this context.
\end{abstract}


\maketitle

\section{Introduction}

It is known that the phase transition do not occur in a finite system, where the thermodynamic functions are always analytic.
However, numerical simulations are usually carried out for finite-size systems.
In order to  study the phase transitions and extract the critical properties from these finite size simulations, one turns to the finite-size scaling (FSS) analysis.
Within the framework of the renormalization group method, finite-size scaling behavior 
can be derived by incorporating the inverse of the system size as a relevant variable \cite{Fisher.1972, Brezin.1985, goldenfeld1992lectures, cardy1996scaling}.
Based on the finite size-scaling ansatz, the critical temperature and the critical exponents can be estimated 
either by collapsing the data  \cite{Stanley.1999, Harada:2011js}, 
or by extrapolating their finite-size estimations to the thermodynamic limit \cite{Shao:2016uw}.
It is hence tempting to directly study systems in the thermodynamic limit, where spontaneous symmetry breaking can occur.
However, it can be challenging to simulate an infinite system.
For example, direct Monte Carlo simulation of a finite domain in an infinite system is highly nontrivial and has only been put forward recently \cite{Herdeiro.2016oqn}.

On the other hand, tensor network (TN) algorithms have emerged recently as an efficient simulation method for the classical models.
It is straightforward to design algorithms which directly simulate infinite-size systems
and a variety of infinite-size tensor network algorithms have been implemented to study the phase transitions of classical models 
\cite{Zhao:2010by, Xie:2012iy, Chen:2017ums,Ueda.2017,Morita.2019,Ueda.2020,Ueda.2021x0n}.
However, it remains challenging to determine accurately the critical properties of the phase transitions.
This is due to the additional length scale induced by the finite cut-off in bond-dimension of the tensor, which is necessary to make the calculation manageable.
This leads to a two length scale scaling problem, which is more difficult to analyze.
For infinite size algorithms it is heuristically known that  the induced length scale scales as a power law in terms of the bond dimension.
One hence can  perform scaling analysis in terms of the bond dimension to circumvent the difficulty of two length scale scaling
\cite{Ueda.2017,Hong:2019cm, Morita.2019, Ueda.2020}.
Conventionally this limit is denoted as the finite-entanglement scaling regime 
since the induced length scale is related to the maximal attainable entanglement in the tensor network.
However, in general one cannot directly tune the induced length scale.
Consequently the scaling analysis in bond dimension is less controlled and 
the analysis might depend on the detail of the tensor renormalization algorithm.

It is also possible to use more sophisticated tensor renormalization algorithm such as the tensor network renormalization (TNR) \cite{Evenbly:2015csa} 
and the loop optimization version of TNR (Loop-TNR) \cite{Yang:2017hj} to obtain the fixed point tensor associated with the phase transition. 
From such a fixed point tensor, the conformal data can be extracted.
However, due to the higher complexity of the algorithm it is more challenging 
to perform the scaling in bond dimension and to generalize the algorithm to higher dimensions.
On the other hand, it has been rare to find tensor network based FSS analysis.
A few exceptions include the studies on two-dimensional (2D) deformed Affleck-Kennedy-Lieb-Tasaki states \cite{Huang.2020wh},
Fisher zeros in the $p$-state clock model \cite{Hong:2019cm} and the Lee–Yang zero of critical XY model \cite{ Hong.2022}.
The positive results in these works indicate that TN based FSS analysis is promising.
How to systematically deal with the crossover from the finite-size scaling regime to the finite-entanglement regime, 
however, remains to be investigated.

In this paper we put forward a scheme which incorporates the finite-size scaling analysis and the tensor network simulation.
We envision the tensor network coarse-graining as a real space renormalization in the tensor space \cite{Efrati:2013tp}.
Starting from the bare tensor which is constructed from the Boltzmann weight of the Hamiltonian, we coarse-grain the tensor iteratively. 
The renormalized tensor is then used to construct the approximated transfer matrix of an infinite strip of finite width.
By diagonalizing the transfer matrix we evaluate relevant physical quantities for the FSS analysis.
Specifically we use higher order tensor renormalization group (HOTRG) \cite{Xie:2012iy} method as our TN coarse-graining engine.
But other tensor renormalization algorithm can also be used in place of the HOTRG.
We benchmark our scheme with 2D classical Ising model and we show that the critical temperature and the critical exponents can be accurately determined.
An important feature of this approach is that the results can be systematically improved by increasing the bond dimension of the HOTRG method.
Furthermore, our results strongly suggest that the crossover between finite-size scaling regime and finite-entanglement scaling regime can be observed.
Based on this, we propose a practical scheme to obtain the best estimation of the critical exponents for a given parameter set.
Finally, the scheme also sheds more light on the nature of the induced length scale in the context of classical models.

The manuscript is organized as follows. 
In Sec.~\ref{sec:model}, we introduce the 2D Ising model and emphasize the relation 
between the partition function, the transfer matrix, and the physical quantities 
such as the free energy density, the correlation length, and the spontaneous magnetization.
We also summarize relevant exact analytical results.
In Sec.~\ref{sec:FSS}, we review the FSS analysis which will be used to estimate the critical temperature and critical exponents.
Then we explain how to evaluate relevant physical quantities within the tensor network framework in Sec.~\ref{sec:TN}.
We present our results in Sec.~\ref{sec:results}, and summarize in Sec.~\ref{sec:summary}.

\section{Model}
\label{sec:model}

We consider 2D classical ferromagnetic Ising model on a square lattice with the Hamiltonian 
\begin{equation}
  H[ \{ s \} ] = -J \sum_{\langle i, j\rangle} s_i s_j - h \sum_i s_i,
\end{equation}
where $J>0$, $s_i=\pm 1$ is the classical spin on site-$i$, $\{ s \}$ denotes the set of all possible spin configurations, 
and $\langle i, j\rangle$ denotes nearest neighbor sites.
It exhibits a high-temperature disordered phase and a low-temperature ferromagnetic ordered phase.
The model is exactly solvable when $h=0$ and the exact critical temperature reads $T_c=2/\ln(1+\sqrt{2})$.
Our numerical study focus on the case of $h=0$.
For a system of $L_x$ columns and $L_y$ rows with the periodic boundary condition,
the partition function can be written as
\begin{equation}
  Z(L_x, L_y) 
  = \sum_{ \{ s \} } e^{-\beta H[\{ s\}]}
  =\text{Tr}\left[ \mathcal{T}^{L_x}(L_y)  \right],
\end{equation}
where $\beta$ is the inverse temperature and 
$\mathcal{T}(L_y)$ is the column-to-column transfer matrix with $L_y$ sites in the column.
Note that $\mathcal{T}(L_y)$ is a $2^{L_y} \times 2^{L_y}$ matrix which can be diagonalized exactly \cite{Kaufman:1949gc, SCHULTZ:1964fv}.
In the following we will denote the eigenvector and the eigenvalue of the transfer matrix as $|\lambda_i(L_y)\rangle$  and $\lambda_i(L_y)$ respectively.
When there is no confusion, we will omit the dependence on $L_y$.

Due to the spin flip symmetry the eigenstates and eigenvalues can be broken into two sectors
and the eigenvalues $\lambda_\pm$ of the even and odd sector can be expressed respectively as
\begin{equation}
  \lambda_{+} = (2\sinh(2K))^{\frac{L_y}{2}} 
  e^{ \frac{1}{2} ( \pm \gamma_1 \pm \gamma_3 \cdots \pm \gamma_{2L_y-1}) }
\end{equation}
and
\begin{equation}
  \lambda_{-} = (2\sinh(2K))^{\frac{L_y}{2}} 
  e^{ \frac{1}{2} ( \pm \gamma_0 \pm \gamma_2 \cdots \pm \gamma_{2L_y-2}) },
\end{equation}
where $K=\beta J$. Here $\gamma_q$ satisfies
\begin{equation}
  \cosh(\gamma_q) = \frac{\cosh(2K)^2}{\sinh(2K)} - \cos\left( \frac{\pi q}{L_y} \right),
\end{equation}
where $q=0, 1, \cdots, 2L_y-1$. 
In particular, the largest and second largest eigenvalues read:
\begin{equation}
  \lambda_{0} = (2\sinh(2K))^{\frac{L_y}{2}} 
  e^{ \frac{1}{2} ( \gamma_1 + \gamma_3 \cdots + \gamma_{2L_y-1}) }
\end{equation}
and
\begin{equation}
  \lambda_{1} = (2\sinh(2K))^{\frac{L_y}{2}} 
  e^{ \frac{1}{2} ( \gamma_0 + \gamma_2 \cdots + \gamma_{2L_y-2}) }.
\end{equation}
When $T \le T_c$, $\lambda_0$ and $\lambda_1$ are degenerate in the thermodynamic limit.

In terms of the $\lambda_i$ the free energy per site reads
\begin{equation}
  -\beta f(L_x, L_y)
  = \frac{ \ln\left( Z(L_x, L_y ) \right)   }{L_x L_y}
  = \frac{ \ln\left( \sum_{i=0}^{2^{L_y}-1} \lambda_i^{L_x}(L_y) \right)   }{L_x L_y}. 
\end{equation}
In the limit of $L_x \rightarrow \infty$ the sum is dominated by the largest eigenvalue, resulting in
\begin{equation}
  -\beta f(L_x=\infty, L_y) = \frac{\ln( \lambda_0(L_y) )}{L_y} = -\frac{E_0(L_y)}{L_y},
\end{equation}
where we define $E_i(L_y) = -\ln \lambda_i(L_y)$.
In other words, for an infinite strip of width $L_y$, the free energy per site is determined by the largest eigenvalue of the transfer matrix.
In the thermodynamic limit the free energy per site $f_\infty$ reads
\begin{equation}
  -\beta f_\infty 
  = \frac{1}{2} \ln\left( 2 \sinh(2K) \right) + \frac{1}{2\pi} \int_0^\pi \gamma(q) dq.
\end{equation}
Furthermore, at $T_c$, the critical free energy per site can be expressed  $-\beta f_\infty(T_c) = \frac{1}{2} \ln(2) + \frac{2}{\pi}G$, where $G=\sum_{n=0}^\infty \frac{(-1)^n}{(2n+1)^2}$ is Catalan's constant.

For an infinite strip, it is straightforward to show that the correlation length is determined by the largest and second-largest eigenvalues of the transfer matrix
\begin{equation}
    \xi(L_x = \infty, L_y) 
    = \frac{1}{\ln \frac{\lambda_0(L_y)}{\lambda_1(L_y)}}
    = \frac{1}{E_1(L_y)-E_0(L_y)}.
\end{equation}
How to evaluate spontaneous magnetization on a finite system, however, is more subtle.
Formally the spontaneous magnetization per site is defined as
\begin{equation}
  m =
  \lim_{h\rightarrow 0} \lim_{L_x, L_y \rightarrow \infty} 
  \left \langle \frac{\sum_i s_i}{N}  \right \rangle.
\end{equation} 
It is important to note that the two limits do not commute.
In the absence of the external field, $m$ is exactly zero as long as the system is finite in any direction.
This is due to the spin-flip symmetry of the Hamiltonian.
To probe the magnetic properties without introducing any external field, 
we define a pseudo spontaneous magnetization as follows:
\begin{equation}
  m_{01} =
  \left| \left\langle \lambda_0 \left| \frac{\sum_i s_i}{N}  \right| \lambda_1 \right\rangle \right|.
\end{equation}
One can show that $m_{01}$ approaches true order parameter $m$ as $L_x, L_y \rightarrow \infty$ \cite{Yang.1952, SCHULTZ:1964fv}.
Furthermore, the energy density can be obtained from the expectation value $-J \langle s_i s_j \rangle$, where $i,j$ are nearest neighbors.
By taking the derivative of the energy density one obtains the specific heat $c_v$.
Finally, from the exact solution of the 2D classical Ising model one can exactly obtain its critical exponents.
The results are: $\nu=1$, $\beta=1/8$, and $\alpha=0$.

\section{Finite-Size Scaling}
\label{sec:FSS}

In this section we describe how to perform FFS analysis to estimate the critical temperature and the critical exponents \cite{Shao:2016uw}.
In the thermodynamic limit, certain physical quantities exhibit power law scaling when the system is near a continuous phase transition.
For example the correlation length $\xi$ scales as
\begin{equation}
  \xi \propto |T-T_c|^{-\nu},
\end{equation}
with the critical exponent $\nu$. 
For a finite-size system, the system size $L$ is the largest length scale
and the scaling above will be replaced by a finite-size scaling form.
In general for a quantity $A$ which scales as $A \propto |T-T_c|^{\kappa}$,
the finite-size scaling form of $A$ reads:
\begin{equation}
  A(T, L) = L^{-\frac{\kappa}{\nu}} f_A((T-T_c) L^{\frac{1}{\nu}}, L^{-\omega}),
\end{equation}
where $f_A$ is the associated scaling function and the term related to $L^{-\omega}$ corresponds to the leading correction to the scaling.
To estimate the critical temperature $T_c$ we first expand the scaling function $f_A$ near $T_c$
\begin{equation}
  A(T, L) L^{\frac{\kappa}{\nu}} =
    a_0 + a_1 t L^{\frac{1}{\nu}} + a_2 t^2 L^{\frac{2}{\nu}} + b_1 L^{-\omega} + c_1 t L^{\frac{1}{\nu}-\omega} \cdots,
\end{equation}
where $t=T-T_c$ and $a_i, b_i, c_i$ are non-universal.
It is then straightforward to show that $A(T, L)$ and $A(T, rL)$ will cross each other at 
\begin{equation}
  \label{eq:FSS_T*}
  T^*_A(L) = T_c 
  + \frac{b_0}{a_0} \frac{ ( 1-r^{- \frac{\kappa}{\nu} } )    }{r^{  \frac{(1-\kappa)}{\nu}} -1 } L^{-\frac{1}{\nu}}
  + \frac{b_1}{a_1} \frac{ ( 1-r^{-\frac{\kappa}{\nu}-\omega} ) }{ r^{ \frac{(1-\kappa)}{\nu}} -1 }  L^{-\frac{1}{\nu}-\omega},
\end{equation}
where $r>1$ is a constant. We shall denote $T^*_A(L)$ as the finite-size critical point and it approaches $T_c$ as $L \rightarrow \infty$.
Note that for dimensionless quantity $Q$ with $\kappa=0$, $T^*_Q(L)$ converges to $T_c$ faster
since the coefficient of the $L^{-1/\nu}$ term is exactly zero.

Critical exponent $\nu$ can be estimated from the slope $s_Q=\frac{dQ}{dT}$ of any dimensionless quantity $Q$.
From the expansion above one finds
\begin{equation}
  s_Q(T, L) = a_1 L^{\frac{1}{\nu}} \left( 
     1 + c_1 L^{-\omega} + 2 a_2 t L^{\frac{1}{\nu}} + \cdots 
  \right).
\end{equation}
At critical point the log of the slope reads
\begin{equation}
  \ln[ s_Q(T_c, L) ] = c + \frac{1}{\nu} \ln(L) + c_1 L^{-\omega} +  \cdots.
\end{equation}
It we take the difference of the log-slope between two sizes $L_1=L$ and $L_2=rL$ at the critical point, one finds
\begin{equation}
  \label{eq:FSS_sQ_Tc}
  \frac{ \ln[s_Q(T_c, rL)] - \ln[s_Q(T_c, L)]}{ \ln(r) } = \frac{1}{\nu} + b L^{-\omega} + \cdots,
\end{equation}
where $b$ is non-universal.
Consequently, the asymptotic behavior of the difference of the log-slope at $T_c$ can be used to estimate $1/\nu$. 
However, it is necessary to first locate the critical point.
Alternatively, one can use the difference of the log-slope at the finite-size critical point $T^*(L)$.
It is straightforward to show that 
\begin{equation}
  \label{eq:FSS_sQ}
  \frac{ \ln[s_Q(T^*(L), rL)] - \ln[s_Q(T^*(L), L)]}{ \ln(r) } = \frac{1}{\nu} + \tilde{b} L^{-\omega} + \cdots,
\end{equation}
with some other non-universal $\tilde{b}$. 
This allows one to extract $1/\nu$ without the knowledge of $T_c$.

Similarly, for dimensionful quantity $A$ with non-zero exponent $\kappa$, one can show that
 the difference of the log-slope at $T_c$ and at $T^*(L)$ scale respectively as
\begin{equation}
  \frac{ \ln[s_A(T_c, rL)] - \ln[s_A(T_c, L)]}{ \ln(r) } = \frac{1-\kappa}{\nu} + b L^{-\omega} + \cdots
\end{equation}
and
\begin{equation}
  \label{eq:FSS_sA_Tc}
  \frac{ \ln[s_AT^*(L), rL)] - \ln[s_A(T^*(L), L)]}{ \ln(r) } = \frac{1-\kappa}{\nu} + \tilde{b} L^{-\omega} + \cdots.
\end{equation}
Hence their asymptotic behavior can be used to extract $(1-\kappa)/\nu$.

\section{Tensor network method}
\label{sec:TN}

\begin{figure}[t]
  \includegraphics[width=0.9\columnwidth]{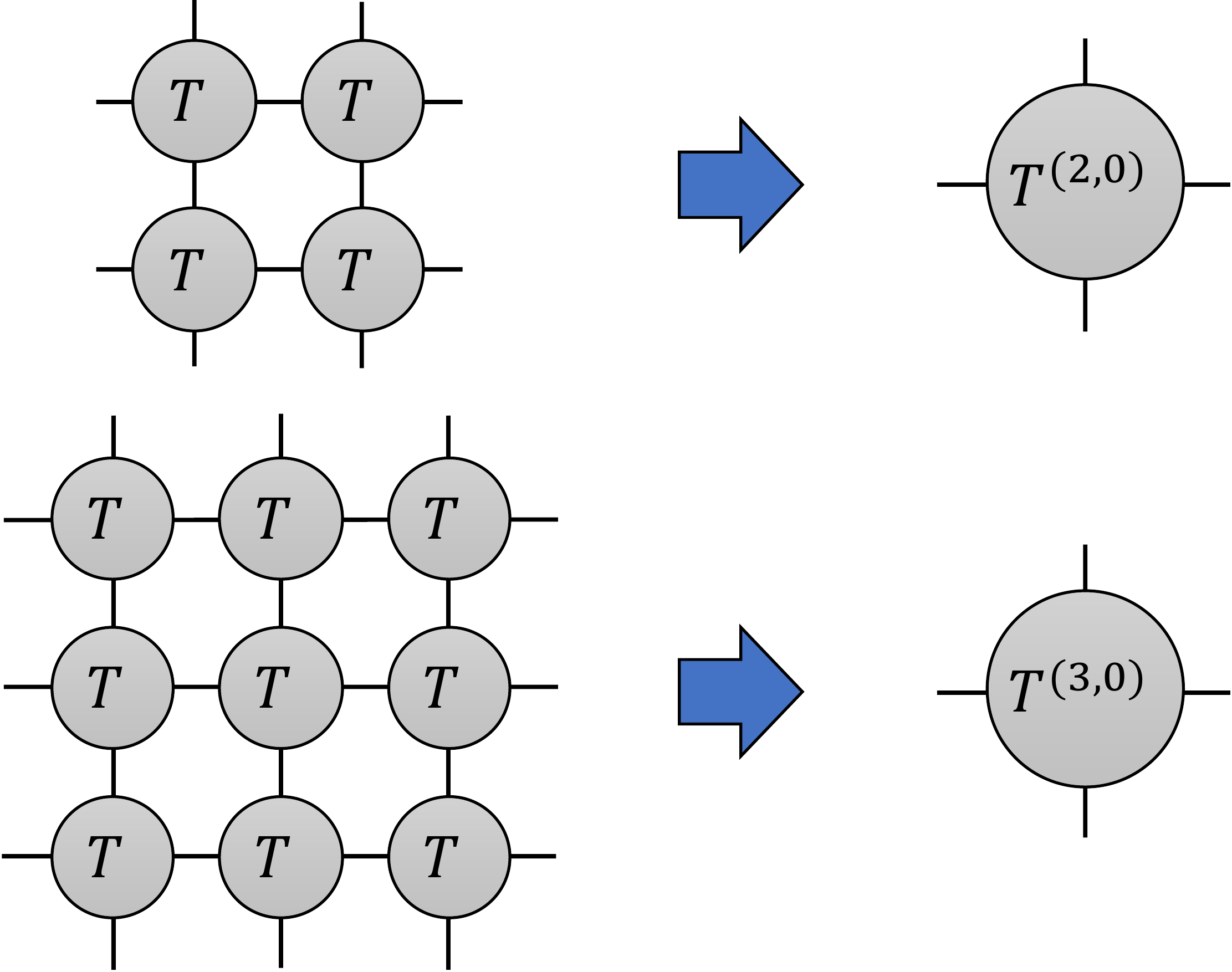}  
  \caption{\label{fig:TN_d}
  Exact tensor contraction of a $2\times 2$ or a $3\times 3$ sites of bare $\mathbf{T}$ tensors into the initial $\mathbf{T}^{(d, 0)}$ tensor.
  }
\end{figure}

\begin{figure}[t]
  \includegraphics[width=0.9\columnwidth]{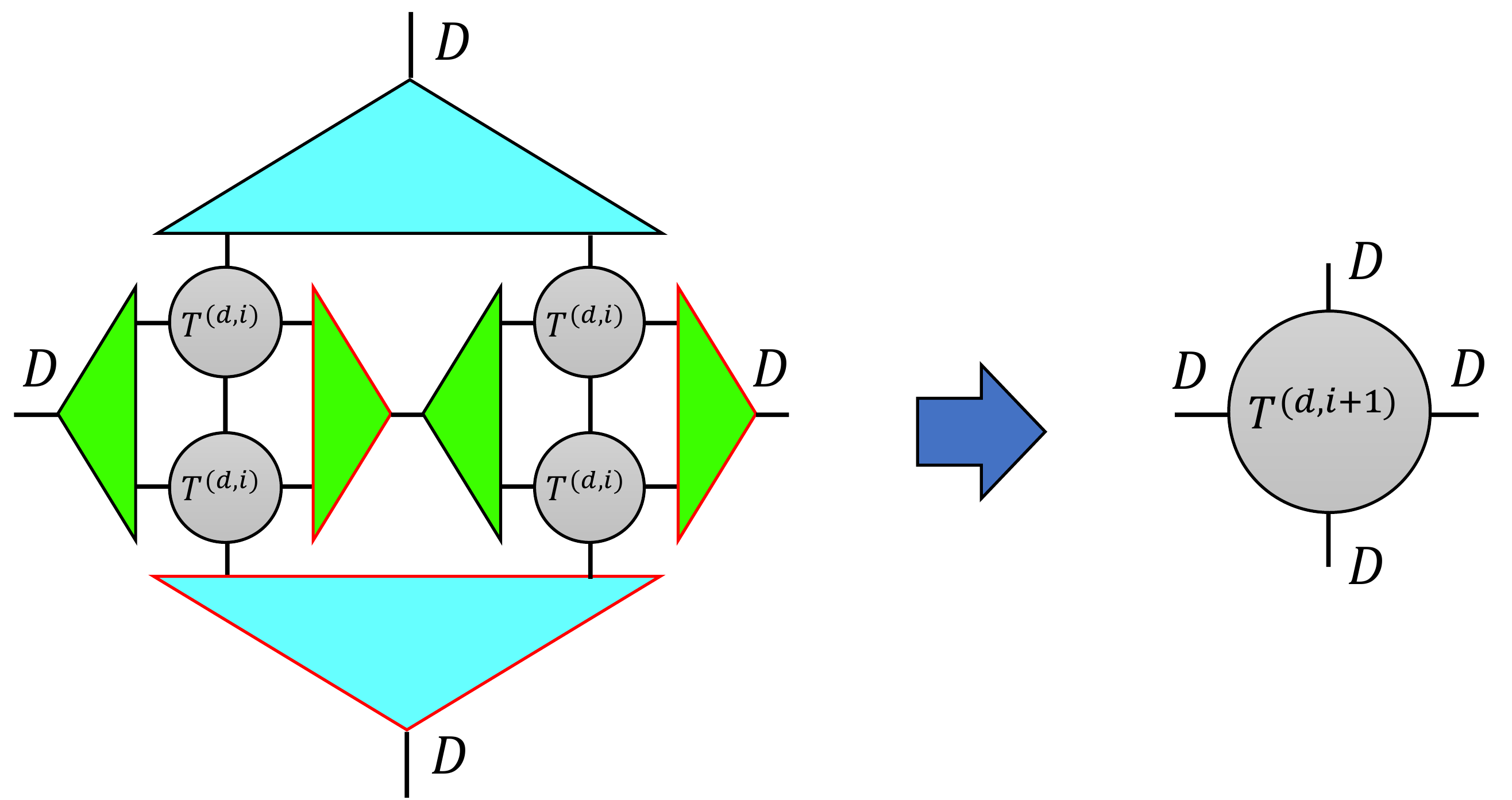}  
  \caption{\label{fig:TN_TTTT} 
  One iteration of HOTRG which renormalizes a $2\times 2$ sites of $\mathbf{T}^{(d, i)}$ tensors into a single $\mathbf{T}^{(d, i+1)}$ tensor.
  }
\end{figure}

In this section we briefly describe how to use tensor network method to evaluate the physical quantities used in the FSS analysis.
It is known that the partition function of the 2D classical Ising model can be written as the tensor trace of a translational invariant tensor network \cite{Xie:2012iy},
\begin{equation}
  Z = \text{tTr} \prod_{\text{sites}} \mathbf{T}_{ijkl},
\end{equation}
where tTr is the tensor trace and the bare tensor $\mathbf{T}_{ijkl}$ is define as
\begin{equation}
  \mathbf{T}_{ijkl}  = \sum_a W_{ai} W_{aj} W_{ak} W_{al},
\end{equation}
and $W$ is determined by the Boltzmann weight on the bond.
In this work we focus on the case of $h=0$ and $W$ reads:
\begin{equation}
  W = \left( \begin{array}{cc}
  \sqrt{ \cosh(\beta J)} & \sqrt{ \sinh(\beta J)} \\
  \sqrt{ \cosh(\beta J)} & -\sqrt{ \sinh(\beta J)} 
  \end{array} \right).
\end{equation}
In 2D, the exact tensor contraction scales exponentially in terms of the system size and many approximated contraction schemes have been proposed.
In this work, we first perform exact tensor contraction of a $d\times d$ sites of bare tensors 
and denote this tensor as $\mathbf{T}^{(d, 0)}$, as sketched in Fig.~\ref{fig:TN_d}.
Specifically, $d=2, 3$ are used to reach more independent sizes.
Then we use the HOTRG \cite{Xie:2012iy} method to renormalize approximately four tensors in a $2\times 2$ geometry 
into a single renoramlized tensor, as sketched in Fig.~\ref{fig:TN_TTTT}.
We denote this as
$\mathbf{T}^{(d, i)} \mathbf{T}^{(d, i)} \mathbf{T}^{(d, i)} \mathbf{T}^{(d, i)} \rightarrow \mathbf{T}^{(d, i+1)}$.
After one HOTRG iteration, the effective size increase by two in both directions.
The accuracy of the HOTRG is controlled by the cut-off bond dimension $D$ in the isometry.

\begin{figure}[t]
  \includegraphics[width=0.9\columnwidth]{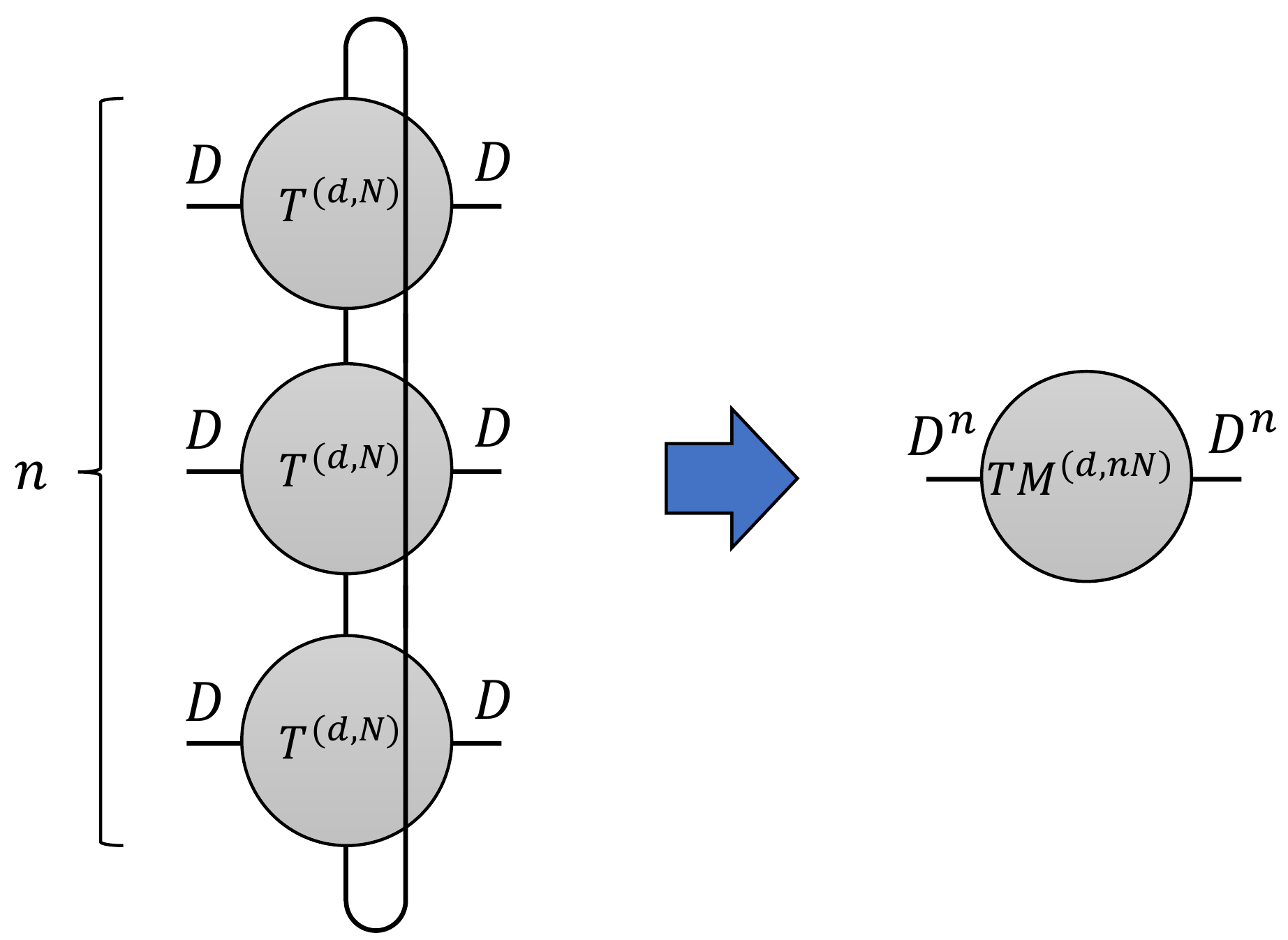}  
  \caption{\label{fig:TN_TM}
  Construction of the column-to-column transfer matrix with $L_y=n*d*2^{N-1}$ and $L_x=d*2^{N-1}$ 
  by stacking $n$ $\mathbf{T}^{(d, N)}$ tensors and tracing out the vertical bonds.
  }
\end{figure}

After $N$ iterations, the renormalized tensor $\mathbf{T}^{(d, N)}$ represents approximately 
a block of of $L \times L$ sites of bare $\mathbf{T}$ tensors, where $L=d *2^{N-1}$.
Now if we trace out the vertical direction then we obtain the approximated column-to-column transfer matrix with $L_y=L$, to the power of $L_x=L$
\begin{equation}
  [\mathcal{T}^{L_x=L}(L_y=L)]_{ab} \approx \mathbf{T}^{(d,N)}_{ab} \equiv \sum_{j} \mathbf{T}^{(d,N)}_{abjj}.
\end{equation}
This results in a $D\times D$ matrix. 
To reach even larger strip width, one can stack $n$ renormalized tensors $\mathbf{T}^{(d, N)}$.
For example with $n=3$ we define a $D^3 \times D^3$ matrix as
\begin{eqnarray}
  & & [\mathcal{T}^{L_x=L}(L_y=3\times L)]_{(a_1a_2a_3),(b_1b_2 b_3)} \nonumber \\
  & \approx  &
  \sum_{j_1 j_2 j_3} \mathbf{T}^{(d,N)}_{a_1 b_1 j_1 j_2} \mathbf{T}^{(d,N)}_{a_2 b_2 j_2 j_3} \mathbf{T}^{(d,N)}_{a_3 b_3 j_3 j_1}.
\end{eqnarray}
This corresponds to the column-to-column transfer matrix with $L_y=3\times L$ sites, to the power of $L_x=L$ as sketched in Fig.~\ref{fig:TN_TM}.
By diagonalizing the approximated transfer matrix (to some power), we obtain the approximated eigenvalues and eigenvectors
from which the free energy per site and the correlation length $\xi$ can be evaluated.
It is also straightforward to evaluate the expectation of the product of local operators such as $\langle s_i s_j \rangle$, where $i, j$ are nearest neighbors.
We refer to Ref.~\cite{Xie:2012iy} for details and only sketch the main idea here.
The first step is to create an impurity tensor $\mathbf{T}^o_{ijkl}$, which depends on the operator $o$. 
Then we replace certain tensors in Fig.~\ref{fig:TN_TTTT}  by the appropriate impurity tensor to obtain the renormalized impurity tensor.
Finally, pseudo spontaneous magnetization $m_{01}$ can be evaluated by contracting the renormalized $\mathbf{T}^z$ tensor 
and the lowest two eigenvectors.

\section{Numerical results}
\label{sec:results}

\subsection{Results based on $\xi/L_y$}

\begin{figure}[t]
  \includegraphics[width=\columnwidth]{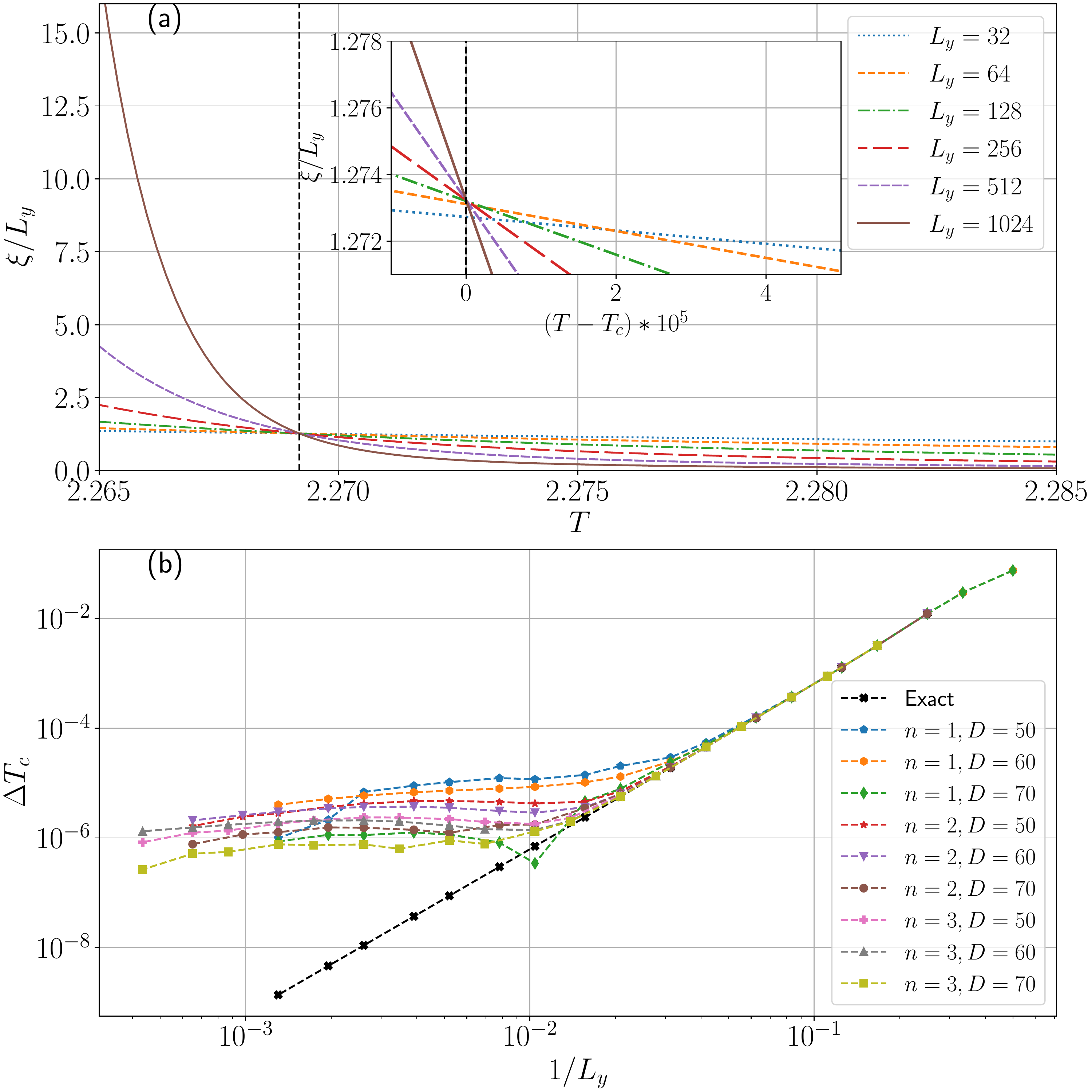}  
  \caption{\label{fig:xi_L} 
  (a) Exact $\xi/L_y$ as a function of the temperature. Vertical dotted line indicates the location of $T_c$. 
  Inset: Zoomed plot near $T_c$.
  (b) Absolute error $\Delta T_c \equiv \left| T_{\xi/L}^*-T_c \right|$ as a function of $1/L_y$.}
\end{figure}

\begin{figure}[t]
  \includegraphics[width=\columnwidth]{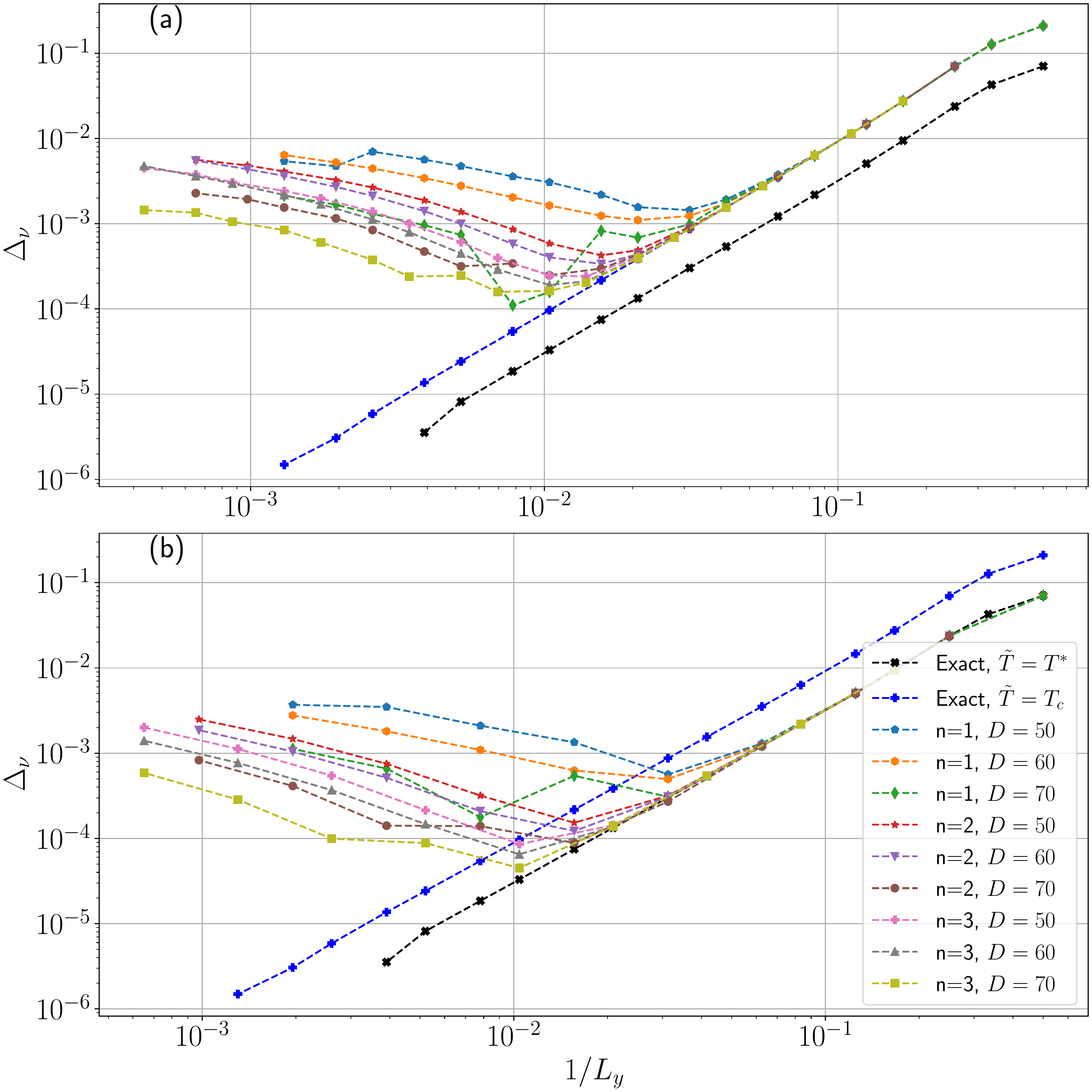}  
  \caption{\label{fig:1_nu} 
  (a) Absolute error $\Delta_\nu$ as a function of $1/L_y$, where the slope of $\xi/L$ at $T_c$ is used.
  (b) Absolute error $\Delta_\nu$ as a function of $1/L_y$, where the slope of $\xi/L$ at $T_{\xi/L}^*(L, 2L)$ is used.}
\end{figure}

We first use the crossing of the dimensionless $\xi/L_y$ to estimate the critical temperature $T_c$.
Here $L_y=n d 2^N$, where $d=2, 3$ is the initial block size, $n=1,2,3$ is the stacking number, and $N$ is the number of iterations.
In Fig.~\ref{fig:xi_L}(a) we plot the exact $\xi/L_y$ as a function of the temperature for various $L_y$.
Here the results are obtained from the exact expression of the $\lambda_0$ and $\lambda_1$.
At this scale, we observe that curves from all sizes cross at $T_c$.
However, if we zoom in near $T_c$ as shown in the inset,
we find that the crossing point deviates from $T_c$, but it drifts toward $T_c$ as $L_y$ increases.
Such a draft is a manifestation of the higher oder corrections to the scaling.
We define finite-size critical temperature $T_{\xi/L_y}^*$ as the crossing point of $\xi/L_y$ for two different sizes $L_y$ and $2L_y$.
In Fig.~\ref{fig:xi_L}(b) we plot the absolute error of the estimated critical temperature 
$\Delta T_c \equiv \left| T_{\xi/L}^*-T_c \right|$ as a function of $1/L_y$ on the log-log scale.
We observe that the exact results (black squares) fall on a straight line, in perfect agreement with Eq.~\ref{eq:FSS_T*}.
The HOTRG results, on the other hand, can be divided into two regimes.
For small $L_y$ the HOTRG results are identical to the exact ones.
At certain length scale, which depends on both $D$ and $n$, they start to deviate from the exact results.
But $\Delta T_c$ continues to decrease as $L_y$ increases.
We hence argue that for a fixed $D$ and $n$ the result at largest size provides the best estimation of the critical temperature.
In Table~\ref{tab:Tc_D}, we summarize our HOTRG results of the absolute error $\Delta T_c$ obtained with $L_y=768$ and various combinations of $D$ and $n$.
It is clear from the table that larger $D$ and $n$ lead to better estimation of the critical temperature.
With $D=70$ one can push the error to the order of $10^{-7}$.

\begin{table}[ht]
  \caption{HOTRG results of the absolute error $\Delta T_c$ obtained with $L_y=768$.}
\begin{center}
\begin{tabular}{|c|c|c|c|c|c|}
  \hline
$D$  &   $n=1$   & $n=2$ & $n=3 $   \\  \hline
  $ 50 $  &  9.09e-7  & 1.64e-6  & 8.34e-7  \\  \hline  
  $ 60 $  &  4.02e-6  & 2.08e-6  & 1.32e-6  \\  \hline    
  $ 70 $  &  8.80e-7  & 8.04e-7  & 3.21e-7  \\  \hline  
\end{tabular}
\end{center}
\label{tab:Tc_D}
\end{table}%

Next we use the slope of the $\xi/L_y$ at $T_c$ or $T_{\xi/L_y}^*$ to estimate $1/\nu$.
Following Eq.~\eqref{eq:FSS_sQ_Tc} and Eq.~\eqref{eq:FSS_sQ} we define 
$\ln \left[ \frac{s_{\xi/L_y}(\tilde{T}_c, 2L_y)}{s_{\xi/L_y}(\tilde{T}_c, L_y)}  \right] \frac{1}{\ln(2)} $ 
as the finite-size estimation of $1/\nu$, where $\tilde{T}_c=T_c$ or $T_{\xi/L_y}^*$.
In Fig.~\ref{fig:1_nu}(a), (b) 
we plot $\Delta_\nu$, the absolute error of the finite size estimation of the $1/\nu$, based respectively on the slope at $T_c$ and $T_{\xi/L_y}^*$.
We find that the exact results (black squares and black diamonds) again fall on a straight line, in perfect agreement with Eqs.~\eqref{eq:FSS_sQ_Tc} and~\eqref{eq:FSS_sQ}.
The difference between two exact results is due to the non-universal prefactor $b$ ($\tilde{b}$) in Eq.~\eqref{eq:FSS_sQ_Tc} (Eq.~\eqref{eq:FSS_sQ}).
On the other hand, the HOTRG results show an interesting crossover behavior. 
For small $L_y$, the HOTRG results follow closely to the exact ones and the error decreases as $L_y$ increases. 
As $L_y$ continues to increase, however, the error will first reach a minima then start to increase. 
We argue that this behavior corresponds to the crossover between the finite-size scaling regime at smaller $L_y$ and the finite-entanglement scaling regime at larger $L_y$.
We hence take the minimal value of finite-size estimation as the best estimation of $1/\nu$ and denote the corresponding system size as $L^*_y$.
It represents the best compromise between the finite-size error and finite-entanglement error for a given $D$ and $n$.
In Table~\ref{tab:nu_D} we summarize our HOTRG results of the absolute error $\Delta_\nu$.
We find again that one can improve the results by increasing $D$ and $n$.
With $D=70$ and $n=3$ one can push the error to the order of $10^{-5}$ or $10^{-4}$
for estimation based on $\tilde{T}_c=T_{\xi/L_y}^*$ at $\tilde{T}_c=T_c$ respectively.

\begin{table}[t]
  \caption{HOTRG results of the absolute error $\Delta_\nu$.}
\begin{center}
\begin{tabular}{|c|c|c|c|c|c|c|c|}
  \hline
  $\tilde{T}=T_c$  &  \multicolumn{2}{|c|}{$n=1$}  & \multicolumn{2}{|c|}{$n=2$}  & \multicolumn{2}{|c|}{$n=3$}   \\  \hline
$D$  &  $L_y^*$  &  $ \Delta_\nu $   & $L_y^*$  & $ \Delta_\nu $ & $L_y^*$  & $ \Delta_\nu  $   \\  \hline
  $ 50 $ & 32   &  1.44e-3  & 64 & 4.25e-4  & 72    & 2.48e-4  \\  \hline  
  $ 60 $ & 48   &  1.10e-3  & 64 & 3.37e-4  & 96    & 1.91e-4  \\  \hline    
  $ 70 $ & 128 &  1.10e-3  & 96 & 2.49e-4  & 144  & 1.58e-4   \\  \hline  
\end{tabular}
\begin{tabular}{|c|c|c|c|c|c|c|c|}
  \hline
  $\tilde{T}=T^*$  &  \multicolumn{2}{|c|}{$n=1$}  & \multicolumn{2}{|c|}{$n=2$}  & \multicolumn{2}{|c|}{$n=3$}   \\  \hline
$D$  &  $L_y^*$  &  $ \Delta_\nu $   & $L_y^*$  & $ \Delta_\nu $ & $L_y^*$  & $ \Delta_\nu $   \\  \hline
  $ 50 $ & 32 &  5.64e-4  & 64 & 1.52e-4  & 96 & 8.58e-5  \\  \hline  
  $ 60 $ & 32 &  4.95e-4  & 64 & 1.22e-4  & 96 & 6.50e-5  \\  \hline    
  $ 70 $ & 128 &  1.78e-4  & 64 & 8.88e-5 & 96  & 4.50e-5   \\  \hline  
\end{tabular}
\end{center}
\label{tab:nu_D}
\end{table}%

\subsection{Results based on pseudo spontaneous magnetization $m_{01}$}

\begin{figure}[t]
  \includegraphics[width=\columnwidth]{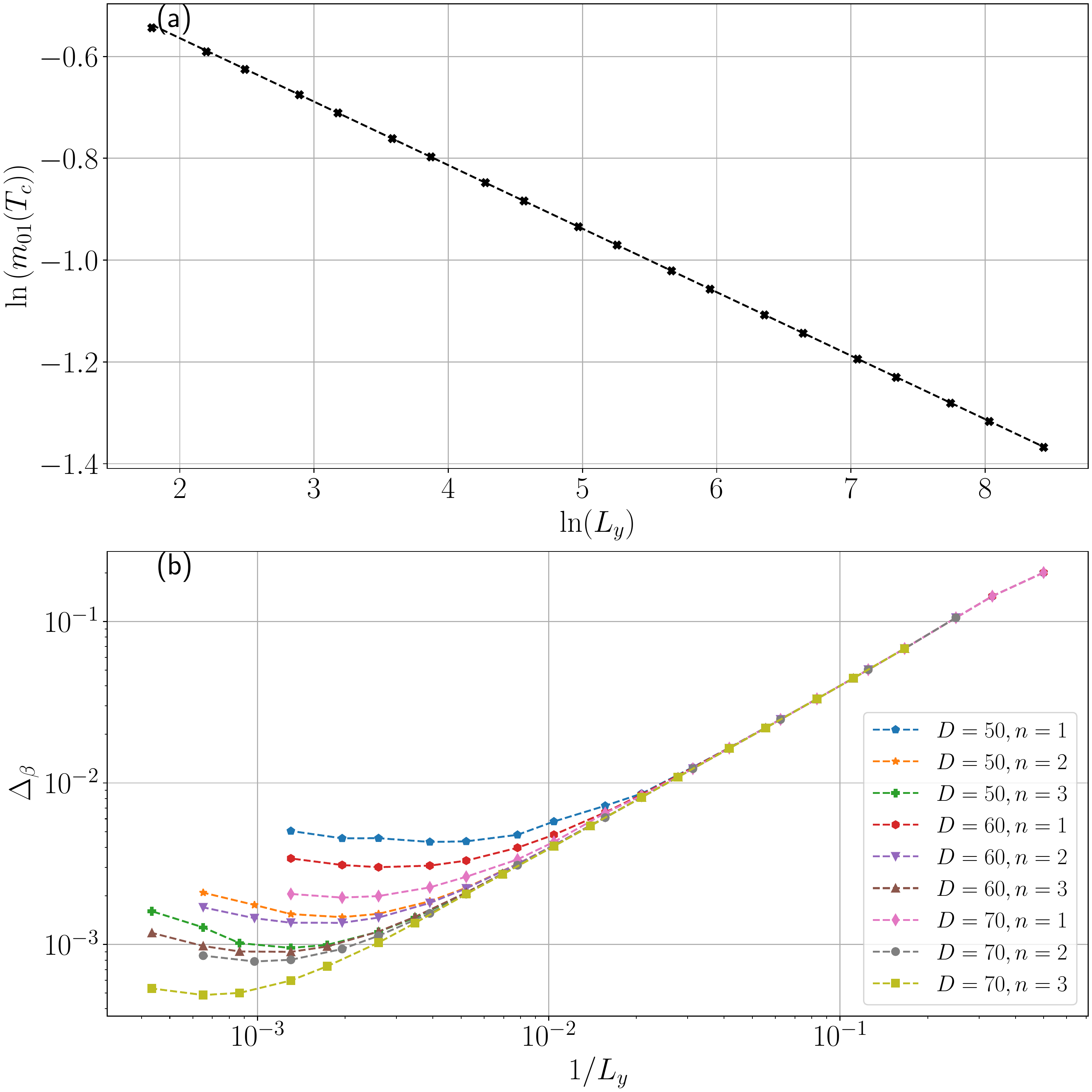}  
  \caption{\label{fig:m01}
  (a) $\ln (m_{01}(T_c))$ as a function of $\ln(L_y)$. Black crosses are the HOTRG results with $D=70$ and $n=3$. 
  Dotted line is the linear fit to the results in the range $16 \le L_y \le 256$.
  (b) Absolute error $\Delta_\beta \equiv \left| \ln \left[ \frac{S_m(T_c, 2L)}{S_m(T_c, L)}  \right]  \frac{1}{\ln(2)} - \frac{1-\beta}{\nu} \right|$ 
  as a function of $1/L_y$, where the slope of $m_{01}$ at $T_c$ is used.}  
\end{figure}

We next use the finite-size scaling behavior of the pseudo spontaneous magnetization $m_{01}$ to estimate the critical exponent $\beta$.
One way to estimate $\beta$ is to study the finite-size scaling of the magnetization at $T_c$.
It is expected that $m_{01}(T_c) \propto L^{-\beta/\nu}$.
In Fig.~\ref{fig:m01}(a) we plot $\ln (m_{01})$ at $T_c$ as a function of $\ln(L_y)$,
where the HOTRG result with $D=70$ and $n=3$ is used. We observe that the results clearly show a linear dependence.
By fitting the results in the range $16 \le L_y \le 256$ we find $\beta/\nu \approx 0.1248$ which is very close to the exact value of $1/8$.
The drawback of this approach is that it might be difficult to determine the range of data to be included in the fitting.
In order to systematically estimate the critical exponent $\beta$ we define $\ln \left[ \frac{S_m(T_c, 2L)}{S_m(T_c, L)}  \right]  \frac{1}{\ln(2)}$ 
as the finite-size estimation of $(1-\beta)/\nu$ according to Eq.~\eqref{eq:FSS_sA_Tc}.
Here we do not use the slope of $m_{01}$ at the finite-size crossing point 
because strictly speaking the definition of $m_{01}$ is meaningful only at $T_c$.
In Fig.~\ref{fig:m01}(b) we plot the absolute error $\Delta_\beta$  as a function of $1/L_y$  on the log-log scale,
and we observe a crossover behavior which is similar to the one in Fig.~\ref{fig:1_nu}.
Based on the same spirit, we take the minimal value of the estimated $(1-\beta)/\nu$ as the best estimation for a given $D$ and $n$.
In Table~\ref{tab:beta} we summarize our results of the absolute error  $\Delta_\beta$.
In general the error is larger than the absolute error $\Delta_\nu$,
but with $D=70$ and $n=3$ one can push the error to the order of $10^{-4}$.

\begin{table}[ht]
  \caption{HOTRG results  of the absolute error $\Delta_\beta$.}
\begin{center}
\begin{tabular}{|c|c|c|c|c|c|c|c|}
  \hline
  &  \multicolumn{2}{|c|}{$n=1$}  & \multicolumn{2}{|c|}{$n=2$}  & \multicolumn{2}{|c|}{$n=3$}   \\  \hline
$D$  &  $L_y^*$  &  $ \Delta_\beta $   & $L_y^*$  & $ \Delta_\beta $ & $L_y^*$  &$ \Delta_\beta $   \\  \hline
  $ 50 $ & 256 &  4.32e-3  & 512 & 1.48e-3  & 768 & 9.53e-4  \\  \hline  
  $ 60 $ & 384 &  3.01e-3  & 512 & 1.36e-4  & 768 & 9.05e-4  \\  \hline    
  $ 70 $ & 512 &  1.95e-3  & 1024 & 7.85e-4 &1536  & 4.87e-4   \\  \hline  
\end{tabular}
\end{center}
\label{tab:beta}
\end{table}%

\subsection{Specific heat $c_v$ and exponent $\alpha$}

\begin{figure}[ht]
  \includegraphics[width=\columnwidth]{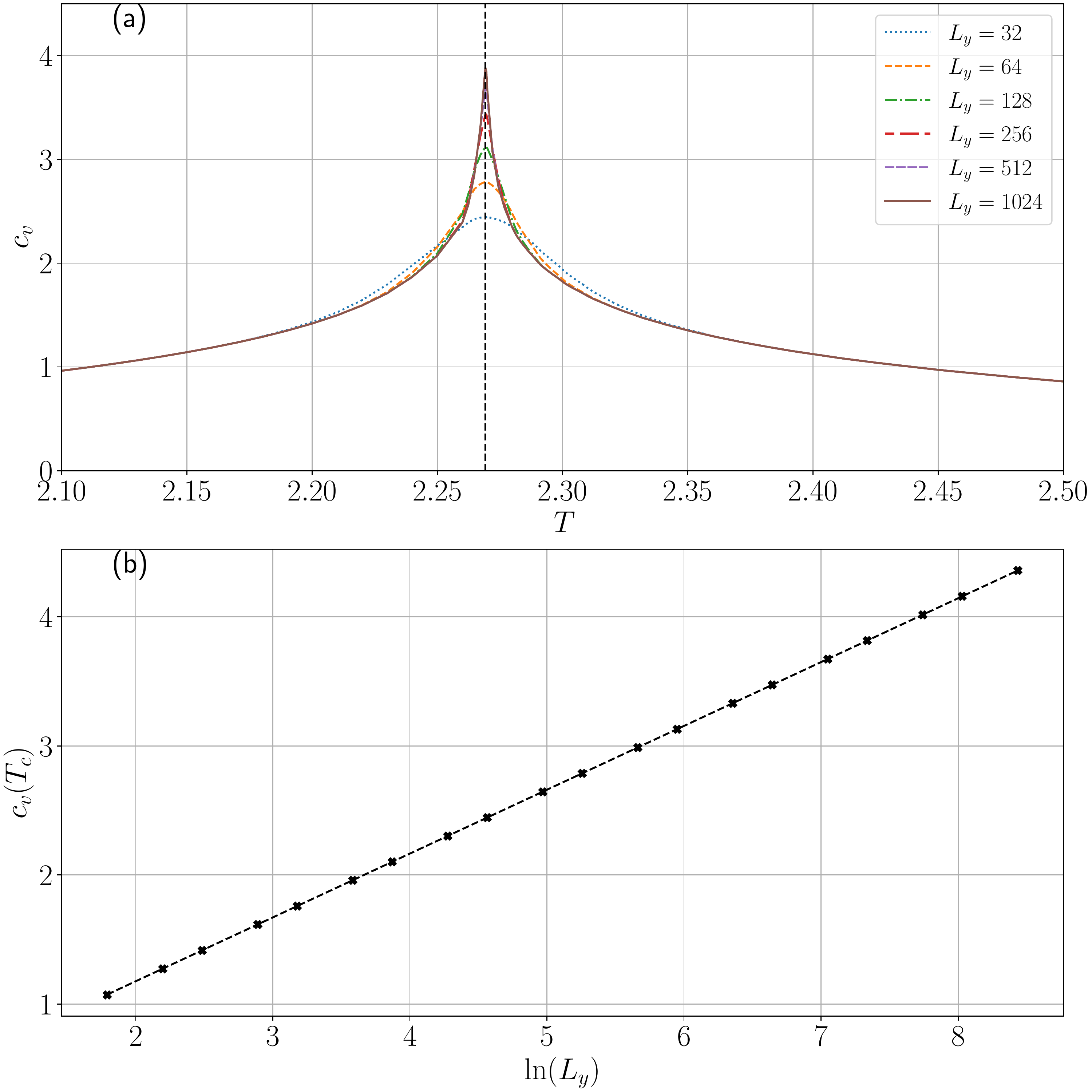}
  \caption{\label{fig:cv} 
  (a) Specific heat as a function of the temperature. 
  (b) Specific heat at $T_c$ as a function of $\ln(L_y)$. Dotted line corresponds to Eq.~\eqref{eq:c_scaling}.}
\end{figure}

It is known that the critical exponent $\alpha$ is zero for the 2D Ising model and the specific heat $c_v $ diverges logarithmically at $T_c$.
Based on the exact solution, it has been shown that for an infinite strip of width $L_y$ the specific heat at $T_c$ scales as \cite{Onsager.1944} 
\begin{eqnarray}
  \label{eq:c_scaling}
  c_v (T_c)
  & \approx  &  \frac{2}{\pi} \left(\ln \cot \frac{\pi}{8} \right)^2 \nonumber 
  \left(\ln L_y + \ln \left( \frac{2^{\frac{5}{2}}}{\pi}\right)+C_E-\frac{\pi}{4} \right) \nonumber \\
  & \approx & 0.4945 \ln L_y+0.1879, 
\end{eqnarray}
where $C_E \approx 0.577215655$ is Euler's constant.
In Fig.~\ref{fig:cv}(a) we plot the HOTRG results with $D=70$ and $n=3$ for the specific heat as a function of the temperature for various systems sizes.
As expected, we observe that it peaks around $T_c$ and the peak becomes higher and shaper as system size increases.
In Fig.~\ref{fig:cv}(b) we plot the specific heat at $T_c$ as a function of $\ln L_y$ and a linear dependence is clearly observed.
Furthermore, the HOTRG results agree excellently with Eq.~\eqref{eq:c_scaling} which is shown as a dotted line.
The excellent agreement between the theory and our results indicates that our method can capture properly the logarithmic divergence of the specific heat and verify that $\alpha=0$.

\subsection{Crossover length scale}
\label{sec:crossover}

\begin{figure}[ht]
  \includegraphics[width=\columnwidth]{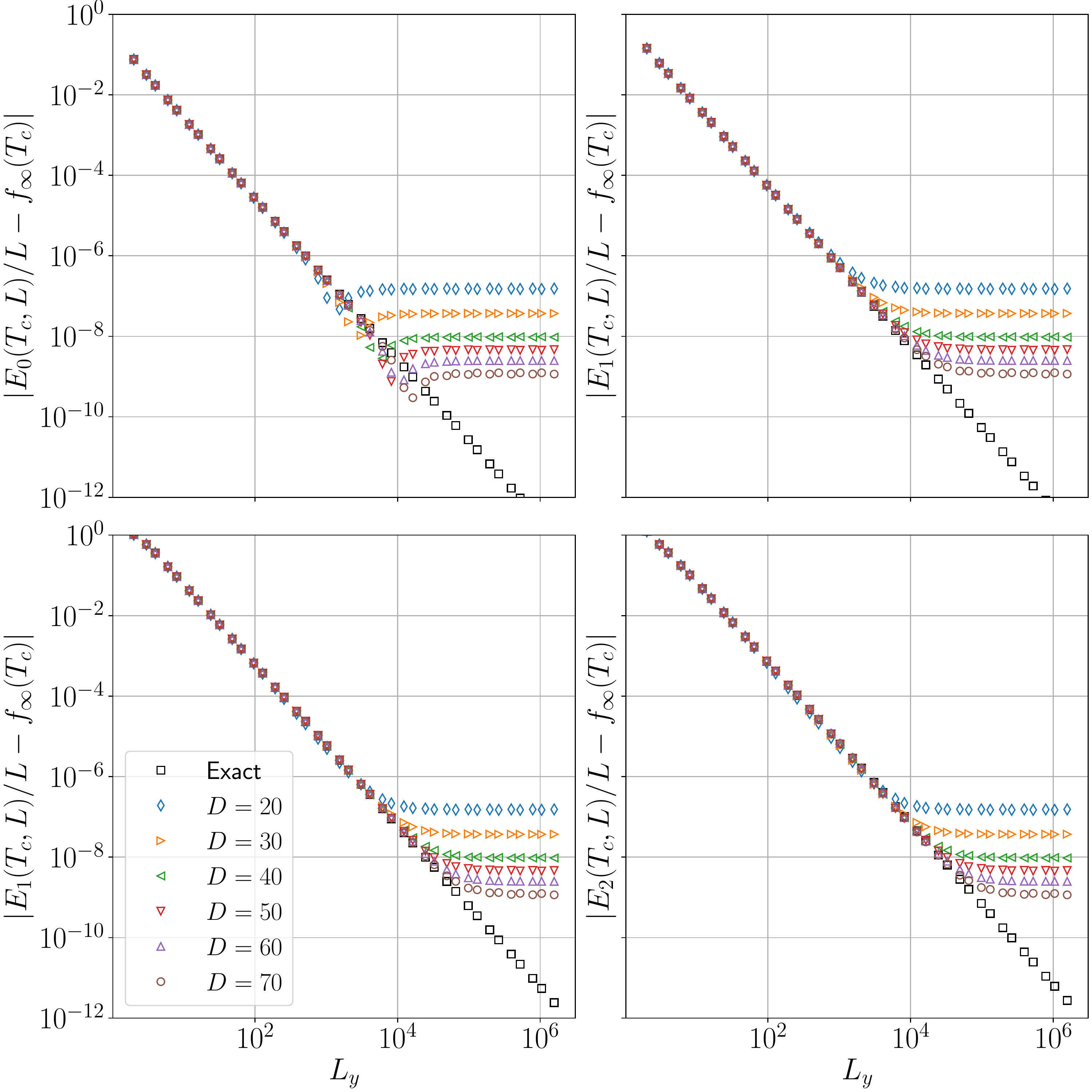}
  \caption{\label{fig:Delta_E} $E_i(T_c, L_y)/L_y - \beta_c f_\infty(T_c)$ as a function of $L_y$ for $i=0,1,2,3$.  
  Black lines corresponds to the exact results. HOTRG results with various $D$ are marked by symbols.}
\end{figure}

The key control parameter of the HOTRG method is the bond-dimension $D$.
It is clear from the results shown about that the accuracy can be systematically improved by increasing $D$.
In general the finite bond dimension in a tensor network calculation gives rise to certain finite length scale.
In the context of the infinite matrix product state for 1D quantum systems, 
it has been shown that at criticality the finite bond dimension induced an effective correlation length
which asymptotically scale as $D^\kappa$ \cite{Pollmann.2009,  Pirvu.2012, Stojevic:2015dj, Vanhecke.2019}.
Here $\kappa$ depends on the central charge $c$ of the underlying conformal field theory through the relation
\begin{equation}
  \label{eq:kappa}
  \kappa = \frac{6}{c\left( \sqrt{\frac{12}{c}} + 1 \right)}.
\end{equation}
Similar scaling behavior has also been observed in the corner-transfer-matrix renormalization group (CTMRG) calculations
for the 2D classical models \cite{Ueda.2014,Ueda.2017,Ueda.2020}. 
We note in passing that abovementioned works focus on the infinite size limit, while in this work we focus on the finite size systems.
It is thus interesting to explore in more detail the nature of the induced length scale in our setting.

\begin{figure}[ht]
  \includegraphics[width=\columnwidth]{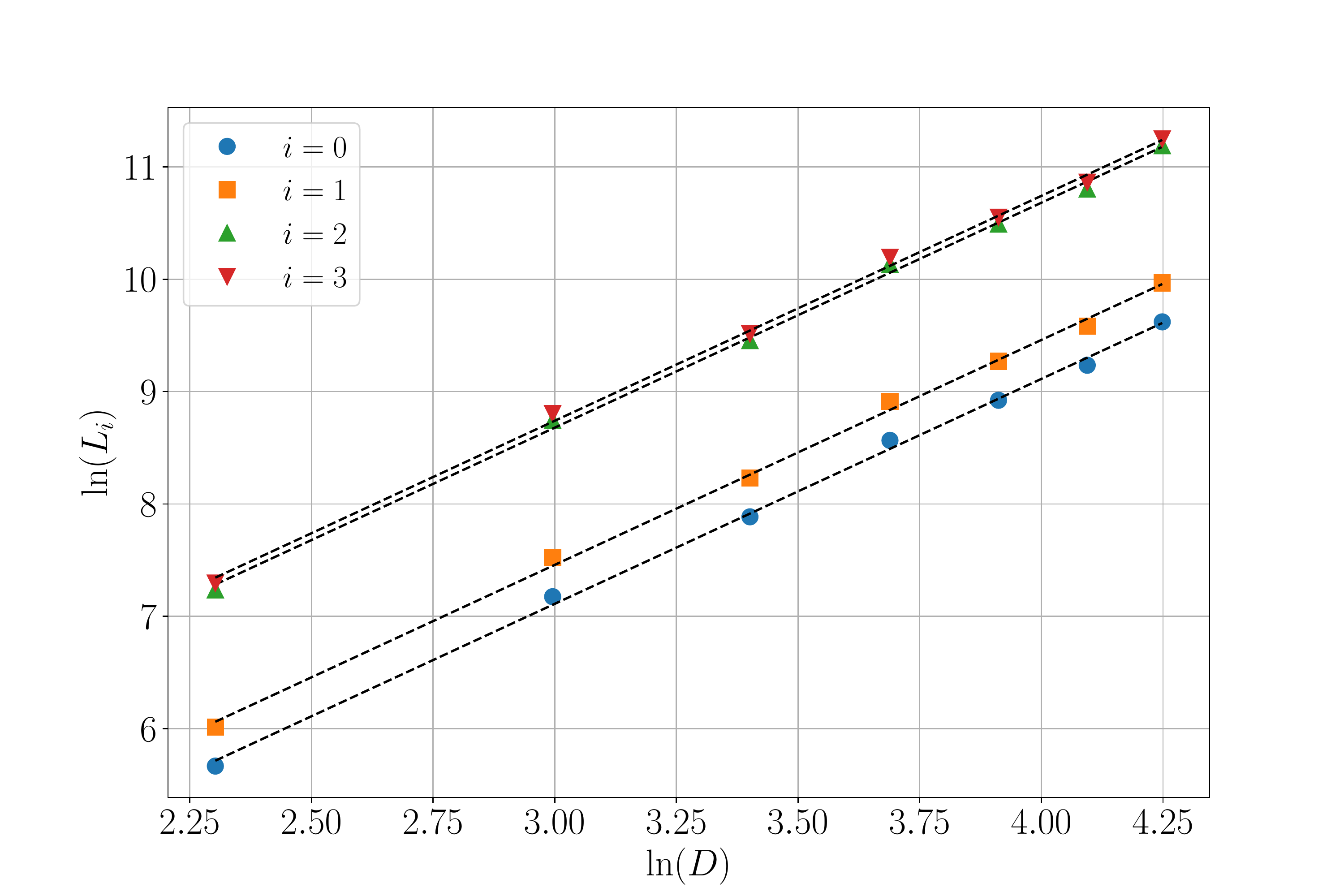}
  \caption{\label{fig:L_cross}  Crossover length scale $L_i$ as a function of $D$.}
\end{figure}

In general at critical temperature one expects that 
\begin{equation}
  \frac{E_i(T_c, L_y)}{L_y} - \beta_c f_\infty(T_c) \propto \frac{1}{L_y^2}.
\end{equation}
In Fig.~\ref{fig:Delta_E} we plot the absolute value of the difference $\left| E_i(T_c, L_y)/L_y - f_\infty(T_c) \right|$ as a function of $L_y$ for $i=0,1,2,3$ on the log-log scale.
We observe that the exact results (black line) fall on the straight lines as expected.
On the other hand, the HOTRG results show different behavior.
First, we find that for a fixed $D$, results from different $n$ collapse into a single curve.
Consequently here we only plots results from $n=1$ for clarity.
Since HOTRG procedure is independent of $n$, this indicates that the induced length scale is solely due to the tensor renormalization.
Furthermore, the results show an interesting crossover behavior.
It is clear from the figure that one can define a crossover length scale $L_i(D)$ for each $i$.
When $L_y < L_i(D)$ the HOTRG results overlap with the exact results, 
while when $L_y > L_i(D)$ the difference saturates to some $i, D$-dependent constant.
We hence interpret $L_i(D)$ as the crossover length scale induced by the finite bond dimension of the HOTRG method.
In Fig.~\ref{fig:L_cross} we plot the crossover length scale $L_i(D)$ as a function of $D$ and a power law dependence is clearly observed.
By fitting the data we find $\kappa \approx 2.001$ for all $i$.
The value is consistent with Eq.~\eqref{eq:kappa}, which gives $\kappa \approx 2.034$ when $c=1/2$ for the 2D classical Ising model is used.

Similar to the matrix product state simulation for one-dimensional quantum systems \cite{Pirvu.2012},
we argue that when $L < L_i(D)$ the system is in the finite-size scaling regime
while when $L > L_i(D)$ the system crosses over to the finite-entanglement scaling regime.
We note that most tensor network calculation for the 2D classical systems work in the infinite size limit,
where the tensor may approach a fixed point tensor and spontaneous symmetry breaking can occur.
These calculations naturally fall into the finite-entanglement regime. 
While in this work, we focus on the finite-size scaling regime.
It also worth mentioning that high accuracy of $E_i$ is needed to accurately determine the critical properties.
Consequently, having $L < L_i(D)$ is not enough to guarantee high accuracy on critical temperature and exponents.
For example, $L_i(D=20)$ is already larger than the system size used in this work, but we find that $D \ge 50$ is necessary to obtain accurate final results.
While in general the exact $E_i$ of the model is not known and one can not following the same procedure to determine $L_i(D)$,
it is still possible to estimate $\kappa$ by fitting $E_i(T_c,D,L\rightarrow \infty)$ as $a + bD^\kappa$.
Here we use the strategy that we increase both $D$ and $n$ and monitor how the estimated quantity change as $1/L$ approaches zero.
This allows to identify the best estimation of the critical temperature or exponents for a fixed $D$ and $n$.
Finally we note that it is possible to use other tensor renormalization scheme instead of the HOTRG to renormalize the $\mathbf{T}$ tensor.
It would be interesting to investigate how the scaling of $L_i(D)$ depends on the specific tensor renormalization scheme.

\section{Summary and discussion}
\label{sec:summary}

In this work we propose a scheme which incorporates tensor network method and finite-size scaling analysis to extract the critical properties of the 2D classical models.
The key ingredient is to use HOTRG method to approximately construct the transfer matrix of an infinite strip of finite width.
By diagonalizing the transfer matrix, physical quantities such as the correlation length and the pseudo magnetization can be straightforwardly obtained.
By applying the finite-size scaling analysis to these quantities, the critical temperature and the critical exponents can be estimated.
An important feature of the scheme is that the results can be systematically improved by tuning the control parameters:
the stacking number $n$ and the bond dimension $D$. 
On the one hand, wider strip width can be reached by increasing $n$. 
This leads to the suppression of the finite-size correction and better estimation of the asymptotic behavior.
On the other hand, larger crossover length can be reached by increasing $D$.
This results in a larger finite-size scaling regime.
Furthermore, we find that the signature of the crossover can be clearly observed
and we propose a strategy to best estimate the critical temperature and the critical exponents for a given $D$ and $n$.

In particular, we use 2D classical Ising model as a benchmark and we show that very accurate results can be obtained.
With $D=70$ and $n=3$, the absolute error of the critical temperature is at the order of $10^{-7}$.
While the absolute error of $1/\nu$ and $(1-\beta)/\nu$ is at the order of $16^{-5}$ and $10^{-4}$ respectively.
We also show the logarithmic divergence of the specific heat and the our results agree excellently with the known analytical expression.
Finally we elucidate the physical meaning of the bond dimension induced length scale. 
We show that for each $E_i$ we can define a crossover length scale $L_i(D)$ which scales as $D^\kappa$.
Furthermore, the value of $\kappa$ agrees with the known equation which is derived from the infinite matrix product state for 1D quantum systems.

We believe that the scheme provides an efficient and systematic way to study the phase transition of the 2D classical models.
As long as the partition function can be expressed as the tensor trace of the a translational invariant tensor network,
one can apply the scheme to locate the critical temperature and to estimate the critical exponents.
While it is expected that for more complex model, larger bond dimension is needed,
one can monitor how the results change as one increases the stacking number $n$ and the bond dimension $D$
and design the strategy to best estimate the critical properties.

Some final comments are now in order.
In this work we focus on isotropic models but it is straightforward to generalize the scheme to anisotropic models.
It is also possible to general the scheme to study three dimensional classical models by using the three dimensional HOTRG.
It also worth mentioning that one can replace the HOTRG method by some other tensor renormalization algorithms.
Since both the accuracy of the renormalized tensor and the computational cost depend on the detail of the algorithm,
it is interesting to investigate which algorithm can provide the best result within the same limit of computational resource.

{\it Note added:}
Near the completion of this work, a close related work \cite{Ueda.2023} appeared. 
In that work, the authors propose a procedure to determine the critical point by combining the finite-size scaling theory of CFT and the numerical data from TNR,
assuming the underlying universality class is known.
In contrast, in this work we focus on how to determine the critical temperature and the critical exponents without knowing the underlying CFT.
They also study the finite bond dimension effect.
Similar to our results in Sec.~\ref{sec:crossover},  an emergence of a finite length which scales as $D^\kappa$ is reported.


\begin{acknowledgments}
We acknowledge the support by National Science and Technology Council (NSTC) of Taiwan through Grant No. 108-2112-M-029-006-MY3, 108-2112-M-002-020-MY3, 110-2112-M-007 -037-MY3.
\end{acknowledgments}

\bibliography{2d_Ising}

\begin{thebibliography}{31}%
\makeatletter
\providecommand \@ifxundefined [1]{%
 \@ifx{#1\undefined}
}%
\providecommand \@ifnum [1]{%
 \ifnum #1\expandafter \@firstoftwo
 \else \expandafter \@secondoftwo
 \fi
}%
\providecommand \@ifx [1]{%
 \ifx #1\expandafter \@firstoftwo
 \else \expandafter \@secondoftwo
 \fi
}%
\providecommand \natexlab [1]{#1}%
\providecommand \enquote  [1]{``#1''}%
\providecommand \bibnamefont  [1]{#1}%
\providecommand \bibfnamefont [1]{#1}%
\providecommand \citenamefont [1]{#1}%
\providecommand \href@noop [0]{\@secondoftwo}%
\providecommand \href [0]{\begingroup \@sanitize@url \@href}%
\providecommand \@href[1]{\@@startlink{#1}\@@href}%
\providecommand \@@href[1]{\endgroup#1\@@endlink}%
\providecommand \@sanitize@url [0]{\catcode `\\12\catcode `\$12\catcode
  `\&12\catcode `\#12\catcode `\^12\catcode `\_12\catcode `\%12\relax}%
\providecommand \@@startlink[1]{}%
\providecommand \@@endlink[0]{}%
\providecommand \url  [0]{\begingroup\@sanitize@url \@url }%
\providecommand \@url [1]{\endgroup\@href {#1}{\urlprefix }}%
\providecommand \urlprefix  [0]{URL }%
\providecommand \Eprint [0]{\href }%
\providecommand \doibase [0]{https://doi.org/}%
\providecommand \selectlanguage [0]{\@gobble}%
\providecommand \bibinfo  [0]{\@secondoftwo}%
\providecommand \bibfield  [0]{\@secondoftwo}%
\providecommand \translation [1]{[#1]}%
\providecommand \BibitemOpen [0]{}%
\providecommand \bibitemStop [0]{}%
\providecommand \bibitemNoStop [0]{.\EOS\space}%
\providecommand \EOS [0]{\spacefactor3000\relax}%
\providecommand \BibitemShut  [1]{\csname bibitem#1\endcsname}%
\let\auto@bib@innerbib\@empty
\bibitem [{\citenamefont {Fisher}\ and\ \citenamefont
  {Barber}(1972)}]{Fisher.1972}%
  \BibitemOpen
  \bibfield  {author} {\bibinfo {author} {\bibfnamefont {M.~E.}\ \bibnamefont
  {Fisher}}\ and\ \bibinfo {author} {\bibfnamefont {M.~N.}\ \bibnamefont
  {Barber}},\ }\bibfield  {title} {\bibinfo {title} {{Scaling Theory for
  Finite-Size Effects in the Critical Region}},\ }\href
  {https://doi.org/10.1103/physrevlett.28.1516} {\bibfield  {journal} {\bibinfo
   {journal} {Phys. Rev. Lett.}\ }\textbf {\bibinfo {volume} {28}},\ \bibinfo
  {pages} {1516} (\bibinfo {year} {1972})}\BibitemShut {NoStop}%
\bibitem [{\citenamefont {Brézin}\ and\ \citenamefont
  {Zinn-Justin}(1985)}]{Brezin.1985}%
  \BibitemOpen
  \bibfield  {author} {\bibinfo {author} {\bibfnamefont {E.}~\bibnamefont
  {Brézin}}\ and\ \bibinfo {author} {\bibfnamefont {J.}~\bibnamefont
  {Zinn-Justin}},\ }\bibfield  {title} {\bibinfo {title} {{Finite size effects
  in phase transitions}},\ }\href
  {https://doi.org/10.1016/0550-3213(85)90379-7} {\bibfield  {journal}
  {\bibinfo  {journal} {Nucl. Phys. B}\ }\textbf {\bibinfo {volume} {257}},\
  \bibinfo {pages} {867} (\bibinfo {year} {1985})}\BibitemShut {NoStop}%
\bibitem [{\citenamefont {Goldenfeld}(1992)}]{goldenfeld1992lectures}%
  \BibitemOpen
  \bibfield  {author} {\bibinfo {author} {\bibfnamefont {N.}~\bibnamefont
  {Goldenfeld}},\ }\href@noop {} {\emph {\bibinfo {title} {{Lectures on Phase
  Transitions and the Renormalization Group}}}},\ Frontiers in physics\
  (\bibinfo  {publisher} {CRC Press, Taylor \& Francis Group},\ \bibinfo {year}
  {1992})\BibitemShut {NoStop}%
\bibitem [{\citenamefont {Cardy}(1996)}]{cardy1996scaling}%
  \BibitemOpen
  \bibfield  {author} {\bibinfo {author} {\bibfnamefont {J.}~\bibnamefont
  {Cardy}},\ }\href@noop {} {\emph {\bibinfo {title} {{Scaling and
  Renormalization in Statistical Physics}}}},\ Cambridge Lecture Notes in
  Physics\ (\bibinfo  {publisher} {Cambridge University Press},\ \bibinfo
  {year} {1996})\BibitemShut {NoStop}%
\bibitem [{\citenamefont {Stanley}(1999)}]{Stanley.1999}%
  \BibitemOpen
  \bibfield  {author} {\bibinfo {author} {\bibfnamefont {H.~E.}\ \bibnamefont
  {Stanley}},\ }\bibfield  {title} {\bibinfo {title} {{Scaling, universality,
  and renormalization: Three pillars of modern critical phenomena}},\ }\href
  {https://doi.org/10.1103/revmodphys.71.s358} {\bibfield  {journal} {\bibinfo
  {journal} {Rev. Mod. Phys.}\ }\textbf {\bibinfo {volume} {71}},\ \bibinfo
  {pages} {S358} (\bibinfo {year} {1999})}\BibitemShut {NoStop}%
\bibitem [{\citenamefont {Harada}(2011)}]{Harada:2011js}%
  \BibitemOpen
  \bibfield  {author} {\bibinfo {author} {\bibfnamefont {K.}~\bibnamefont
  {Harada}},\ }\bibfield  {title} {{\selectlanguage {English}\bibinfo {title}
  {{Bayesian inference in the scaling analysis of critical phenomena}}},\
  }\href {https://doi.org/10.1103/physreve.84.056704} {\bibfield  {journal}
  {\bibinfo  {journal} {Phys. Rev. E}\ }\textbf {\bibinfo {volume} {84}},\
  \bibinfo {pages} {056704} (\bibinfo {year} {2011})}\BibitemShut {NoStop}%
\bibitem [{\citenamefont {Shao}\ \emph {et~al.}(2016)\citenamefont {Shao},
  \citenamefont {Guo},\ and\ \citenamefont {Sandvik}}]{Shao:2016uw}%
  \BibitemOpen
  \bibfield  {author} {\bibinfo {author} {\bibfnamefont {H.}~\bibnamefont
  {Shao}}, \bibinfo {author} {\bibfnamefont {W.}~\bibnamefont {Guo}},\ and\
  \bibinfo {author} {\bibfnamefont {A.~W.}\ \bibnamefont {Sandvik}},\
  }\bibfield  {title} {\bibinfo {title} {{Quantum criticality with two length
  scales}},\ }\href {https://doi.org/10.1126/science.aad5007} {\bibfield
  {journal} {\bibinfo  {journal} {Science}\ }\textbf {\bibinfo {volume}
  {352}},\ \bibinfo {pages} {213} (\bibinfo {year} {2016})}\BibitemShut
  {NoStop}%
\bibitem [{\citenamefont {Herdeiro}\ and\ \citenamefont
  {Doyon}(2016)}]{Herdeiro.2016oqn}%
  \BibitemOpen
  \bibfield  {author} {\bibinfo {author} {\bibfnamefont {V.}~\bibnamefont
  {Herdeiro}}\ and\ \bibinfo {author} {\bibfnamefont {B.}~\bibnamefont
  {Doyon}},\ }\bibfield  {title} {\bibinfo {title} {{Monte Carlo method for
  critical systems in infinite volume: The planar Ising model}},\ }\href
  {https://doi.org/10.1103/physreve.94.043322} {\bibfield  {journal} {\bibinfo
  {journal} {Phys. Rev. E}\ }\textbf {\bibinfo {volume} {94}},\ \bibinfo
  {pages} {043322} (\bibinfo {year} {2016})}\BibitemShut {NoStop}%
\bibitem [{\citenamefont {Zhao}\ \emph {et~al.}(2010)\citenamefont {Zhao},
  \citenamefont {Xie}, \citenamefont {Chen}, \citenamefont {Wei}, \citenamefont
  {Cai},\ and\ \citenamefont {Xiang}}]{Zhao:2010by}%
  \BibitemOpen
  \bibfield  {author} {\bibinfo {author} {\bibfnamefont {H.~H.}\ \bibnamefont
  {Zhao}}, \bibinfo {author} {\bibfnamefont {Z.~Y.}\ \bibnamefont {Xie}},
  \bibinfo {author} {\bibfnamefont {Q.~N.}\ \bibnamefont {Chen}}, \bibinfo
  {author} {\bibfnamefont {Z.~C.}\ \bibnamefont {Wei}}, \bibinfo {author}
  {\bibfnamefont {J.~W.}\ \bibnamefont {Cai}},\ and\ \bibinfo {author}
  {\bibfnamefont {T.}~\bibnamefont {Xiang}},\ }\bibfield  {title}
  {{\selectlanguage {English}\bibinfo {title} {{Renormalization of
  tensor-network states}}},\ }\href
  {https://doi.org/10.1103/physrevb.81.174411} {\bibfield  {journal} {\bibinfo
  {journal} {Phys. Rev. B}\ }\textbf {\bibinfo {volume} {81}},\ \bibinfo
  {pages} {174411} (\bibinfo {year} {2010})}\BibitemShut {NoStop}%
\bibitem [{\citenamefont {Xie}\ \emph {et~al.}(2012)\citenamefont {Xie},
  \citenamefont {Chen}, \citenamefont {Qin}, \citenamefont {Zhu}, \citenamefont
  {Yang},\ and\ \citenamefont {Xiang}}]{Xie:2012iy}%
  \BibitemOpen
  \bibfield  {author} {\bibinfo {author} {\bibfnamefont {Z.-Y.}\ \bibnamefont
  {Xie}}, \bibinfo {author} {\bibfnamefont {J.}~\bibnamefont {Chen}}, \bibinfo
  {author} {\bibfnamefont {M.-P.}\ \bibnamefont {Qin}}, \bibinfo {author}
  {\bibfnamefont {J.~W.}\ \bibnamefont {Zhu}}, \bibinfo {author} {\bibfnamefont
  {L.~P.}\ \bibnamefont {Yang}},\ and\ \bibinfo {author} {\bibfnamefont
  {T.}~\bibnamefont {Xiang}},\ }\bibfield  {title} {{\selectlanguage
  {English}\bibinfo {title} {{Coarse-graining renormalization by higher-order
  singular value decomposition}}},\ }\href
  {https://doi.org/10.1103/physrevb.86.045139} {\bibfield  {journal} {\bibinfo
  {journal} {Phys. Rev. B}\ }\textbf {\bibinfo {volume} {86}},\ \bibinfo
  {pages} {045139} (\bibinfo {year} {2012})}\BibitemShut {NoStop}%
\bibitem [{\citenamefont {Chen}\ \emph {et~al.}(2017)\citenamefont {Chen},
  \citenamefont {Liao}, \citenamefont {Xie}, \citenamefont {Han}, \citenamefont
  {Huang}, \citenamefont {Cheng}, \citenamefont {Wei}, \citenamefont {Xie},\
  and\ \citenamefont {Xiang}}]{Chen:2017ums}%
  \BibitemOpen
  \bibfield  {author} {\bibinfo {author} {\bibfnamefont {J.}~\bibnamefont
  {Chen}}, \bibinfo {author} {\bibfnamefont {H.-J.}\ \bibnamefont {Liao}},
  \bibinfo {author} {\bibfnamefont {H.-D.}\ \bibnamefont {Xie}}, \bibinfo
  {author} {\bibfnamefont {X.-J.}\ \bibnamefont {Han}}, \bibinfo {author}
  {\bibfnamefont {R.-Z.}\ \bibnamefont {Huang}}, \bibinfo {author}
  {\bibfnamefont {S.}~\bibnamefont {Cheng}}, \bibinfo {author} {\bibfnamefont
  {Z.-C.}\ \bibnamefont {Wei}}, \bibinfo {author} {\bibfnamefont {Z.-Y.}\
  \bibnamefont {Xie}},\ and\ \bibinfo {author} {\bibfnamefont {T.}~\bibnamefont
  {Xiang}},\ }\bibfield  {title} {\bibinfo {title} {{Phase Transition of the
  q-State Clock Model: Duality and Tensor Renormalization}},\ }\href
  {https://doi.org/10.1088/0256-307x/34/5/050503} {\bibfield  {journal}
  {\bibinfo  {journal} {Chin. Phys. Lett.}\ }\textbf {\bibinfo {volume} {34}},\
  \bibinfo {pages} {050503} (\bibinfo {year} {2017})}\BibitemShut {NoStop}%
\bibitem [{\citenamefont {Ueda}\ \emph {et~al.}(2017)\citenamefont {Ueda},
  \citenamefont {Okunishi}, \citenamefont {Krčmár}, \citenamefont {Gendiar},
  \citenamefont {Yunoki},\ and\ \citenamefont {Nishino}}]{Ueda.2017}%
  \BibitemOpen
  \bibfield  {author} {\bibinfo {author} {\bibfnamefont {H.}~\bibnamefont
  {Ueda}}, \bibinfo {author} {\bibfnamefont {K.}~\bibnamefont {Okunishi}},
  \bibinfo {author} {\bibfnamefont {R.}~\bibnamefont {Krčmár}}, \bibinfo
  {author} {\bibfnamefont {A.}~\bibnamefont {Gendiar}}, \bibinfo {author}
  {\bibfnamefont {S.}~\bibnamefont {Yunoki}},\ and\ \bibinfo {author}
  {\bibfnamefont {T.}~\bibnamefont {Nishino}},\ }\bibfield  {title} {\bibinfo
  {title} {{Critical behavior of the two-dimensional icosahedron model}},\
  }\href {https://doi.org/10.1103/physreve.96.062112} {\bibfield  {journal}
  {\bibinfo  {journal} {Phys. Rev. E}\ }\textbf {\bibinfo {volume} {96}},\
  \bibinfo {pages} {062112} (\bibinfo {year} {2017})}\BibitemShut {NoStop}%
\bibitem [{\citenamefont {Morita}\ and\ \citenamefont
  {Kawashima}(2019)}]{Morita.2019}%
  \BibitemOpen
  \bibfield  {author} {\bibinfo {author} {\bibfnamefont {S.}~\bibnamefont
  {Morita}}\ and\ \bibinfo {author} {\bibfnamefont {N.}~\bibnamefont
  {Kawashima}},\ }\bibfield  {title} {{\selectlanguage {English}\bibinfo
  {title} {{Calculation of higher-order moments by higher-order tensor
  renormalization group}}},\ }\href {https://doi.org/10.1016/j.cpc.2018.10.014}
  {\bibfield  {journal} {\bibinfo  {journal} {Comput. Phys. Commun.}\ }\textbf
  {\bibinfo {volume} {236}},\ \bibinfo {pages} {65} (\bibinfo {year}
  {2019})}\BibitemShut {NoStop}%
\bibitem [{\citenamefont {Ueda}\ \emph {et~al.}(2020)\citenamefont {Ueda},
  \citenamefont {Okunishi}, \citenamefont {Harada}, \citenamefont {Krčmár},
  \citenamefont {Gendiar}, \citenamefont {Yunoki},\ and\ \citenamefont
  {Nishino}}]{Ueda.2020}%
  \BibitemOpen
  \bibfield  {author} {\bibinfo {author} {\bibfnamefont {H.}~\bibnamefont
  {Ueda}}, \bibinfo {author} {\bibfnamefont {K.}~\bibnamefont {Okunishi}},
  \bibinfo {author} {\bibfnamefont {K.}~\bibnamefont {Harada}}, \bibinfo
  {author} {\bibfnamefont {R.}~\bibnamefont {Krčmár}}, \bibinfo {author}
  {\bibfnamefont {A.}~\bibnamefont {Gendiar}}, \bibinfo {author} {\bibfnamefont
  {S.}~\bibnamefont {Yunoki}},\ and\ \bibinfo {author} {\bibfnamefont
  {T.}~\bibnamefont {Nishino}},\ }\bibfield  {title} {{\selectlanguage
  {English}\bibinfo {title} {{Finite-m scaling analysis of
  Berezinskii-Kosterlitz-Thouless phase transitions and entanglement spectrum
  for the six-state clock model}}},\ }\href
  {https://doi.org/10.1103/physreve.101.062111} {\bibfield  {journal} {\bibinfo
   {journal} {Phys. Rev. E}\ }\textbf {\bibinfo {volume} {101}},\ \bibinfo
  {pages} {062111} (\bibinfo {year} {2020})}\BibitemShut {NoStop}%
\bibitem [{\citenamefont {Ueda}\ and\ \citenamefont
  {Oshikawa}(2021)}]{Ueda.2021x0n}%
  \BibitemOpen
  \bibfield  {author} {\bibinfo {author} {\bibfnamefont {A.}~\bibnamefont
  {Ueda}}\ and\ \bibinfo {author} {\bibfnamefont {M.}~\bibnamefont
  {Oshikawa}},\ }\bibfield  {title} {\bibinfo {title} {{Resolving the
  Berezinskii-Kosterlitz-Thouless transition in the two-dimensional XY model
  with tensor-network-based level spectroscopy}},\ }\href
  {https://doi.org/10.1103/physrevb.104.165132} {\bibfield  {journal} {\bibinfo
   {journal} {Phys. Rev. B}\ }\textbf {\bibinfo {volume} {104}},\ \bibinfo
  {pages} {165132} (\bibinfo {year} {2021})}\BibitemShut {NoStop}%
\bibitem [{\citenamefont {Hong}\ and\ \citenamefont {Kim}(2020)}]{Hong:2019cm}%
  \BibitemOpen
  \bibfield  {author} {\bibinfo {author} {\bibfnamefont {S.}~\bibnamefont
  {Hong}}\ and\ \bibinfo {author} {\bibfnamefont {D.-H.}\ \bibnamefont {Kim}},\
  }\bibfield  {title} {{\selectlanguage {English}\bibinfo {title} {{Logarithmic
  finite-size scaling correction to the leading Fisher zeros in the p-state
  clock model: A higher-order tensor renormalization group study}}},\ }\href
  {https://doi.org/10.1103/physreve.101.012124} {\bibfield  {journal} {\bibinfo
   {journal} {Phys. Rev. E}\ }\textbf {\bibinfo {volume} {101}},\ \bibinfo
  {pages} {012124} (\bibinfo {year} {2020})}\BibitemShut {NoStop}%
\bibitem [{\citenamefont {Evenbly}\ and\ \citenamefont
  {Vidal}(2015)}]{Evenbly:2015csa}%
  \BibitemOpen
  \bibfield  {author} {\bibinfo {author} {\bibfnamefont {G.}~\bibnamefont
  {Evenbly}}\ and\ \bibinfo {author} {\bibfnamefont {G.}~\bibnamefont
  {Vidal}},\ }\bibfield  {title} {{\selectlanguage {English}\bibinfo {title}
  {{Tensor Network Renormalization}}},\ }\href
  {https://doi.org/10.1103/physrevlett.115.180405} {\bibfield  {journal}
  {\bibinfo  {journal} {Phys. Rev. Lett.}\ }\textbf {\bibinfo {volume} {115}},\
  \bibinfo {pages} {180405} (\bibinfo {year} {2015})}\BibitemShut {NoStop}%
\bibitem [{\citenamefont {Yang}\ \emph {et~al.}(2017)\citenamefont {Yang},
  \citenamefont {Gu},\ and\ \citenamefont {Wen}}]{Yang:2017hj}%
  \BibitemOpen
  \bibfield  {author} {\bibinfo {author} {\bibfnamefont {S.}~\bibnamefont
  {Yang}}, \bibinfo {author} {\bibfnamefont {Z.-C.}\ \bibnamefont {Gu}},\ and\
  \bibinfo {author} {\bibfnamefont {X.-G.}\ \bibnamefont {Wen}},\ }\bibfield
  {title} {{\selectlanguage {English}\bibinfo {title} {{Loop Optimization for
  Tensor Network Renormalization}}},\ }\href
  {https://doi.org/10.1103/physrevlett.118.110504} {\bibfield  {journal}
  {\bibinfo  {journal} {Phys. Rev. Lett.}\ }\textbf {\bibinfo {volume} {118}},\
  \bibinfo {pages} {110504} (\bibinfo {year} {2017})}\BibitemShut {NoStop}%
\bibitem [{\citenamefont {Huang}\ \emph {et~al.}(2020)\citenamefont {Huang},
  \citenamefont {Lu},\ and\ \citenamefont {Chen}}]{Huang.2020wh}%
  \BibitemOpen
  \bibfield  {author} {\bibinfo {author} {\bibfnamefont {C.-Y.}\ \bibnamefont
  {Huang}}, \bibinfo {author} {\bibfnamefont {Y.-C.}\ \bibnamefont {Lu}},\ and\
  \bibinfo {author} {\bibfnamefont {P.}~\bibnamefont {Chen}},\ }\bibfield
  {title} {\bibinfo {title} {{Finite-size scaling analysis of two-dimensional
  deformed Affleck-Kennedy-Lieb-Tasaki states}},\ }\href
  {https://doi.org/10.1103/physrevb.102.165108} {\bibfield  {journal} {\bibinfo
   {journal} {Phys. Rev. B}\ }\textbf {\bibinfo {volume} {102}},\ \bibinfo
  {pages} {165108} (\bibinfo {year} {2020})}\BibitemShut {NoStop}%
\bibitem [{\citenamefont {Hong}\ and\ \citenamefont {Kim}(2022)}]{Hong.2022}%
  \BibitemOpen
  \bibfield  {author} {\bibinfo {author} {\bibfnamefont {S.}~\bibnamefont
  {Hong}}\ and\ \bibinfo {author} {\bibfnamefont {D.-H.}\ \bibnamefont {Kim}},\
  }\bibfield  {title} {\bibinfo {title} {{Tensor Network Calculation of the
  Logarithmic Correction Exponent in the XY Model}},\ }\href
  {https://doi.org/10.7566/jpsj.91.084003} {\bibfield  {journal} {\bibinfo
  {journal} {JPSJ}\ }\textbf {\bibinfo {volume} {91}},\ \bibinfo {pages}
  {084003} (\bibinfo {year} {2022})}\BibitemShut {NoStop}%
\bibitem [{\citenamefont {Efrati}\ \emph {et~al.}(2013)\citenamefont {Efrati},
  \citenamefont {Wang}, \citenamefont {Kolan},\ and\ \citenamefont
  {Kadanoff}}]{Efrati:2013tp}%
  \BibitemOpen
  \bibfield  {author} {\bibinfo {author} {\bibfnamefont {E.}~\bibnamefont
  {Efrati}}, \bibinfo {author} {\bibfnamefont {Z.}~\bibnamefont {Wang}},
  \bibinfo {author} {\bibfnamefont {A.}~\bibnamefont {Kolan}},\ and\ \bibinfo
  {author} {\bibfnamefont {L.~P.}\ \bibnamefont {Kadanoff}},\ }\bibfield
  {title} {\bibinfo {title} {{Real-space renormalization in statistical
  mechanics}},\ }\href {https://doi.org/10.1103/revmodphys.86.647} {\bibfield
  {journal} {\bibinfo  {journal} {Rev. Mod. Phys.}\ }\textbf {\bibinfo {volume}
  {86}},\ \bibinfo {pages} {647} (\bibinfo {year} {2013})}\BibitemShut
  {NoStop}%
\bibitem [{\citenamefont {Kaufman}(1949)}]{Kaufman:1949gc}%
  \BibitemOpen
  \bibfield  {author} {\bibinfo {author} {\bibfnamefont {B.}~\bibnamefont
  {Kaufman}},\ }\bibfield  {title} {{\selectlanguage {English}\bibinfo {title}
  {{Crystal Statistics. II. Partition Function Evaluated by Spinor
  Analysis}}},\ }\href {https://doi.org/10.1103/physrev.76.1232} {\bibfield
  {journal} {\bibinfo  {journal} {Phys. Rev.}\ }\textbf {\bibinfo {volume}
  {76}},\ \bibinfo {pages} {1232 } (\bibinfo {year} {1949})}\BibitemShut
  {NoStop}%
\bibitem [{\citenamefont {SCHULTZ}\ \emph {et~al.}(1964)\citenamefont
  {SCHULTZ}, \citenamefont {Mattis},\ and\ \citenamefont
  {Lieb}}]{SCHULTZ:1964fv}%
  \BibitemOpen
  \bibfield  {author} {\bibinfo {author} {\bibfnamefont {T.~D.}\ \bibnamefont
  {SCHULTZ}}, \bibinfo {author} {\bibfnamefont {D.}~\bibnamefont {Mattis}},\
  and\ \bibinfo {author} {\bibfnamefont {E.~H.}\ \bibnamefont {Lieb}},\
  }\bibfield  {title} {{\selectlanguage {English}\bibinfo {title}
  {{Two-Dimensional Ising Model as a Soluble Problem of Many Fermions}}},\
  }\href {https://doi.org/10.1103/revmodphys.36.856} {\bibfield  {journal}
  {\bibinfo  {journal} {Rev. Mod. Phys.}\ }\textbf {\bibinfo {volume} {36}},\
  \bibinfo {pages} {856 } (\bibinfo {year} {1964})}\BibitemShut {NoStop}%
\bibitem [{\citenamefont {Yang}(1952)}]{Yang.1952}%
  \BibitemOpen
  \bibfield  {author} {\bibinfo {author} {\bibfnamefont {C.~N.}\ \bibnamefont
  {Yang}},\ }\bibfield  {title} {\bibinfo {title} {{The Spontaneous
  Magnetization of a Two-Dimensional Ising Model}},\ }\href
  {https://doi.org/10.1103/physrev.85.808} {\bibfield  {journal} {\bibinfo
  {journal} {Phys. Rev.}\ }\textbf {\bibinfo {volume} {85}},\ \bibinfo {pages}
  {808} (\bibinfo {year} {1952})}\BibitemShut {NoStop}%
\bibitem [{\citenamefont {Onsager}(1944)}]{Onsager.1944}%
  \BibitemOpen
  \bibfield  {author} {\bibinfo {author} {\bibfnamefont {L.}~\bibnamefont
  {Onsager}},\ }\bibfield  {title} {\bibinfo {title} {{Crystal Statistics. I. A
  Two-Dimensional Model with an Order-Disorder Transition}},\ }\href
  {https://doi.org/10.1103/physrev.65.117} {\bibfield  {journal} {\bibinfo
  {journal} {Phys. Rev.}\ }\textbf {\bibinfo {volume} {65}},\ \bibinfo {pages}
  {117} (\bibinfo {year} {1944})}\BibitemShut {NoStop}%
\bibitem [{\citenamefont {Pollmann}\ \emph {et~al.}(9 06)\citenamefont
  {Pollmann}, \citenamefont {Mukerjee}, \citenamefont {Turner},\ and\
  \citenamefont {Moore}}]{Pollmann.2009}%
  \BibitemOpen
  \bibfield  {author} {\bibinfo {author} {\bibfnamefont {F.}~\bibnamefont
  {Pollmann}}, \bibinfo {author} {\bibfnamefont {S.}~\bibnamefont {Mukerjee}},
  \bibinfo {author} {\bibfnamefont {A.~M.}\ \bibnamefont {Turner}},\ and\
  \bibinfo {author} {\bibfnamefont {J.~E.}\ \bibnamefont {Moore}},\ }\bibfield
  {title} {{\selectlanguage {English}\bibinfo {title} {{Theory of
  Finite-Entanglement Scaling at One-Dimensional Quantum Critical Points}}},\
  }\href {https://doi.org/10.1103/physrevlett.102.255701} {\bibfield  {journal}
  {\bibinfo  {journal} {Phys. Rev. Lett.}\ }\textbf {\bibinfo {volume} {102}},\
  \bibinfo {pages} {255701} (\bibinfo {year} {2009-06})}\BibitemShut {NoStop}%
\bibitem [{\citenamefont {Pirvu}\ \emph {et~al.}(2012)\citenamefont {Pirvu},
  \citenamefont {Vidal}, \citenamefont {Verstraete},\ and\ \citenamefont
  {Tagliacozzo}}]{Pirvu.2012}%
  \BibitemOpen
  \bibfield  {author} {\bibinfo {author} {\bibfnamefont {B.}~\bibnamefont
  {Pirvu}}, \bibinfo {author} {\bibfnamefont {G.}~\bibnamefont {Vidal}},
  \bibinfo {author} {\bibfnamefont {F.}~\bibnamefont {Verstraete}},\ and\
  \bibinfo {author} {\bibfnamefont {L.}~\bibnamefont {Tagliacozzo}},\
  }\bibfield  {title} {\bibinfo {title} {{Matrix product states for critical
  spin chains: Finite-size versus finite-entanglement scaling}},\ }\href
  {https://doi.org/10.1103/physrevb.86.075117} {\bibfield  {journal} {\bibinfo
  {journal} {Phys. Rev. B}\ }\textbf {\bibinfo {volume} {86}},\ \bibinfo
  {pages} {075117} (\bibinfo {year} {2012})}\BibitemShut {NoStop}%
\bibitem [{\citenamefont {Stojevic}\ \emph {et~al.}(2015)\citenamefont
  {Stojevic}, \citenamefont {Haegeman}, \citenamefont {McCulloch},
  \citenamefont {Tagliacozzo},\ and\ \citenamefont
  {Verstraete}}]{Stojevic:2015dj}%
  \BibitemOpen
  \bibfield  {author} {\bibinfo {author} {\bibfnamefont {V.}~\bibnamefont
  {Stojevic}}, \bibinfo {author} {\bibfnamefont {J.}~\bibnamefont {Haegeman}},
  \bibinfo {author} {\bibfnamefont {I.~P.}\ \bibnamefont {McCulloch}}, \bibinfo
  {author} {\bibfnamefont {L.}~\bibnamefont {Tagliacozzo}},\ and\ \bibinfo
  {author} {\bibfnamefont {F.}~\bibnamefont {Verstraete}},\ }\bibfield  {title}
  {{\selectlanguage {English}\bibinfo {title} {{Conformal data from finite
  entanglement scaling}}},\ }\href {https://doi.org/10.1103/physrevb.91.035120}
  {\bibfield  {journal} {\bibinfo  {journal} {Phys. Rev. B}\ }\textbf {\bibinfo
  {volume} {91}},\ \bibinfo {pages} {035120} (\bibinfo {year}
  {2015})}\BibitemShut {NoStop}%
\bibitem [{\citenamefont {Vanhecke}\ \emph {et~al.}(2019)\citenamefont
  {Vanhecke}, \citenamefont {Haegeman}, \citenamefont {Acoleyen}, \citenamefont
  {Vanderstraeten},\ and\ \citenamefont {Verstraete}}]{Vanhecke.2019}%
  \BibitemOpen
  \bibfield  {author} {\bibinfo {author} {\bibfnamefont {B.}~\bibnamefont
  {Vanhecke}}, \bibinfo {author} {\bibfnamefont {J.}~\bibnamefont {Haegeman}},
  \bibinfo {author} {\bibfnamefont {K.~V.}\ \bibnamefont {Acoleyen}}, \bibinfo
  {author} {\bibfnamefont {L.}~\bibnamefont {Vanderstraeten}},\ and\ \bibinfo
  {author} {\bibfnamefont {F.}~\bibnamefont {Verstraete}},\ }\bibfield  {title}
  {\bibinfo {title} {{Scaling Hypothesis for Matrix Product States}},\ }\href
  {https://doi.org/10.1103/physrevlett.123.250604} {\bibfield  {journal}
  {\bibinfo  {journal} {Phys. Rev. Lett.}\ }\textbf {\bibinfo {volume} {123}},\
  \bibinfo {pages} {250604} (\bibinfo {year} {2019})}\BibitemShut {NoStop}%
\bibitem [{\citenamefont {Ueda}\ \emph {et~al.}(2014)\citenamefont {Ueda},
  \citenamefont {Okunishi},\ and\ \citenamefont {Nishino}}]{Ueda.2014}%
  \BibitemOpen
  \bibfield  {author} {\bibinfo {author} {\bibfnamefont {H.}~\bibnamefont
  {Ueda}}, \bibinfo {author} {\bibfnamefont {K.}~\bibnamefont {Okunishi}},\
  and\ \bibinfo {author} {\bibfnamefont {T.}~\bibnamefont {Nishino}},\
  }\bibfield  {title} {\bibinfo {title} {{Doubling of entanglement spectrum in
  tensor renormalization group}},\ }\href
  {https://doi.org/10.1103/physrevb.89.075116} {\bibfield  {journal} {\bibinfo
  {journal} {Phys. Rev. B}\ }\textbf {\bibinfo {volume} {89}},\ \bibinfo
  {pages} {075116} (\bibinfo {year} {2014})}\BibitemShut {NoStop}%
\bibitem [{\citenamefont {Ueda}\ and\ \citenamefont {Oshikawa}()}]{Ueda.2023}%
  \BibitemOpen
  \bibfield  {author} {\bibinfo {author} {\bibfnamefont {A.}~\bibnamefont
  {Ueda}}\ and\ \bibinfo {author} {\bibfnamefont {M.}~\bibnamefont
  {Oshikawa}},\ }\bibfield  {title} {\bibinfo {title} {{Finite-size and finite
  bond dimension effects of tensor network renormalization}},\ }\href@noop {}
  {\bibfield  {journal} {\bibinfo  {journal} {arXiv}\ }}\Eprint
  {https://arxiv.org/abs/2302.06632} {2302.06632} \BibitemShut {NoStop}%
\end{thebibliography}%

\end{document}